\documentclass[%
 reprint,
 amsmath,amssymb,
 aps,
]{revtex4-2}
\usepackage{xcolor} 
\usepackage{graphicx}% Include figure files
\usepackage{dcolumn}% Align table columns on decimal point
\usepackage{bm}% bold math
\begin{document}

\preprint{APS/123-QED}

\title{A probabilistic approach to system-environment coupling}
\author{Mohammad Rahbar}
\affiliation{Technical University of Munich; TUM School of Natural Sciences, Department of Chemistry, Lichtenbergstr. 4, D-85748 Garching, Germany}
\author{Christopher J. Stein}%
\affiliation{Technical University of Munich; TUM School of Natural Sciences, Department of Chemistry, Catalysis Research Center, Atomistic Modeling Center, Munich Data Science Institute,Lichtenbergstr. 4, D-85748 Garching, Germany}
\email{christopher.stein@tum.de}

\date{\today}% It is always \today, today,

\begin{abstract}
We introduce a unified statistical framework for quantifying system–environment coupling by treating the interaction energy $V_\mathcal{SE}$ as a stochastic variable. Using a reference‑particle decomposition, we derive exact, closed‑form expressions for the mean and variance of $V_\mathcal{SE}$ in terms of the single‑particle density and up to four‑body correlation functions. When $V_\mathcal{SE}$ is approximately Gaussian, these two moments suffice to compute the free energy shift of the strongly coupled system. To validate our framework, we ran explicit Monte Carlo simulations of the full system–environment configurations across a range of system sizes, generating reference distributions of the interaction energy $V_\mathcal{SE}$. We then applied our derived analytical formulas to predict these distributions and found excellent agreement in both the weak‑ and strong‑coupling regimes.
\end{abstract}

\maketitle

%\tableofcontents

\section{Introduction}

The accurate evaluation of free energy differences \cite{kirkwood1935statistical,widom1963some,torrie1977nonphysical,bennett1976efficient,kastner2011umbrella,henin2022enhanced,chipot2007free} and excess chemical potential \cite{pohorille1996excess,beck2006potential,widom1982potential,bansal2017quasichemical,paliwal2006analysis} remains a challenging problem \cite{mitchell1991free} in statistical mechanics, particularly when strong coupling between a system and its environment renders traditional partition-function-based techniques ineffective \cite{ashbaugh2006colloquium,jarzynski2017stochastic,vilar2008failure,PhysRevLett.101.098901,PhysRevLett.101.098903,seifert2012stochastic,van2015ensemble,esposito2010entropy,seifert2016first,talkner2016open,miller2017entropy,goldstein2019nonequilibrium,campa2009statistical,jarzynski2017stochastic,binder2018thermodynamics,talkner2020colloquium,thirring2003negative,hanggi1990reaction,ochoa2016energy,xing2024thermodynamics,ding2022strong,wang2012achieving}.
In such regimes, the interaction energy between the system and its surroundings, denoted by $V_\mathcal{SE}$ plays an important role \cite{talkner2016open,talkner2020colloquium,jarzynski2017stochastic,campisi2009fluctuation,PRL} since both the excess chemical potential and free energy changes can be expressed as functionals of the probability distribution of $V_\mathcal{SE}$\cite{chipot2007free,beck2006potential,asthagiri2003absolute,mey2020best,york2023modern}.
Under conditions where $V_\mathcal{SE}$ is nearly Gaussian, its impact on thermodynamic potentials reduces to contributions from the first two statistical moments \cite{miao2014improved,wang2024convergence,wang2021gaussian,miao2015gaussian}. 
In the present work, we develop a comprehensive statistical mechanical framework to characterize the interaction energy $V_{SE}$ as a stochastic variable. 
Focusing on an equilibrium system composed of two subsystems --- the system $\mathcal{S}$ and its environment $\mathcal{E}$ --- we derive exact expressions for the expectation $\langle V_\mathcal{SE} \rangle$ and the variance $\mathrm{Var}(V_\mathcal{SE})$. 
This derivation is accomplished using our proposed reference-particle decomposition, which provides a systematic  way to capture many-body correlations.
While we do not directly compute thermodynamic potentials in this paper, the quantities $\langle V_\mathcal{SE} \rangle$ and $\mathrm{Var}(V_\mathcal{SE})$ establish the statistical foundation necessary for such calculations. 
As an example, in a dilute solution, the excess chemical potential of inserting a single solute molecule of type $\xi$ can be written using the Potential Distribution Theorem \cite{jackson1964potential,widom1963some,widom1982potential,beck2006potential,pratt2001quasi,lawrence1998quasi,pratt1999quasi,paulaitis2002hydration,kirkwood1935statistical,azimi2025potential} as
\begin{equation}
\mu_{\xi}^{\mathrm{ex}} = -\beta^{-1} \ln \, \langle\langle e^{-\beta V_{SE}} \rangle\rangle_0,
\end{equation}
where $V_\mathcal{SE}$ denotes the interaction energy between the solute and the surrounding solvent. 
The double brackets $\langle\langle \cdots \rangle\rangle_0$ represent an average over an ensemble in which the solute and the solvent are thermally equilibrated but non-interacting, meaning that the solute is present but does not perturb the environment \cite{beck2006potential}. 
This expression corresponds exactly to the free energy difference between two thermodynamic states: the reference state in which the solute and solvent are non-interacting, and the fully interacting state in which the solute is coupled to the environment through $V_\mathcal{SE}$. 
When $V_\mathcal{SE}$ is approximately Gaussian distributed, this implies \cite{zwanzig1954high,chipot2007free,beck2006potential,kubo1962generalized,widom1963some,wang2024convergence}
\begin{equation}
\mu_\xi^{\mathrm{ex}} \approx \langle\langle V_{SE} \rangle\rangle_0 - \frac{\beta\, \mathrm{Var}_0(V_{SE})}{2}.
\end{equation}
Thus, the precise determination of the first and second moments provides a practical route to evaluate thermodynamic quantities associated with system–environment coupling. 
In our accompanying article \cite{PRL}, we show that the probability distribution of the interaction energy, $P(V_\mathcal{SE})$, serves as a central object in defining thermodynamic potentials for strongly coupled open systems under specific conditions.
In particular, when $P(V_\mathcal{SE})$ is approximately Gaussian, accurate knowledge of its mean and variance becomes essential for quantifying how the equilibrium probability distribution function $P_\beta(x_\mathcal{S})$ of the system—when it is coupled to the environment deviates from, and is reshaped relative to, its uncoupled counterpart $P(x_\mathcal{S})$. 
The derivations presented here are  model-independent under the sole assumption of pairwise interactions.
In the remainder of the article, we present the derivation of the first and second moments of $V_{\mathcal{SE}}$ via a reference-particle approach and discuss the asymptotic behavior of the interaction energy distribution in large systems.
In Section \ref{sec2} we define the interaction energy $V_{\mathcal{SE}}$ and its distribution.
Section \ref{rpa} introduces the reference‑particle decomposition approach we leverage to obtain $\langle V_{\mathcal{SE}}\rangle$ and $\langle V_{\mathcal{SE}}^2\rangle$, which are explicitly derived in Sections \ref{DFM} and \ref{DSM}, respectively. 
In Section \ref{secfree}, we extend the framework to use these moments to compute the free energy shift $\Delta F_S$ and finally validate our predictions via Monte Carlo simulations in Section \ref{secmode} before we conclude with a summary and outlook in the final Section \ref{seccon}.

\section{Definition of system-environment interactions}
\label{sec2}
\begin{figure}[htbp]
    \centering
    \includegraphics[width=\columnwidth]{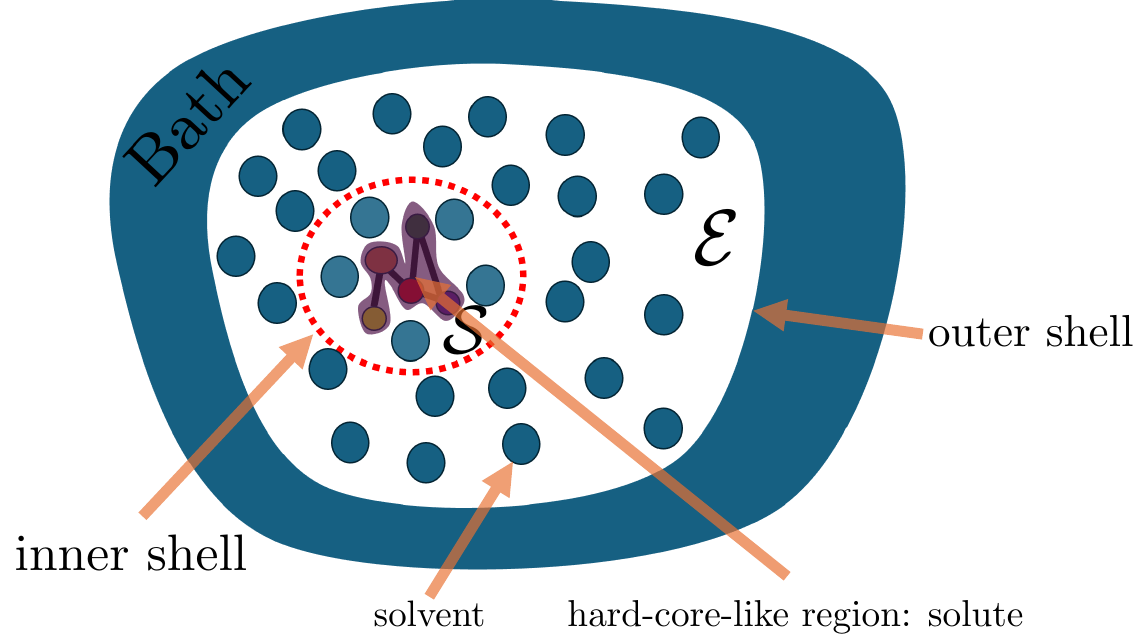}
    \caption{Solvation of a dilute molecular solution as an example of a strongly-coupled open system to introduce the notation for the system-environment definition used throughout this work.}
    \label{solution}
\end{figure}

Here, we strictly define the interaction energy as used throughout this work. 
As a concrete example that helps illustrating our approach, we consider a dilute molecular solution in which a single solute molecule (e.g. a protein) is immersed in a large number of solvent molecules (e.g. water).
In such a scenario, it is natural to define $\mathcal{S}$ as the inner shell--- comprising solvent molecules in close proximity to the solute --- and $\mathcal{E}$ as the outer shell, which contains the remaining, more weakly interacting solvents \cite{gomez2022hydrated,rempe2004inner,asthagiri2003absolute,beck2006potential}. 
In this work, we develop a general framework for characterizing the interaction between a system and its environment.
For notational simplicity, we assume that the only non-negligible interaction arises between solvent particles located inside $\mathcal{S}$ and those in $\mathcal{E}$.
The solute itself is treated as a hard-core-like region that contributes solely through excluded volume effects. 
It imposes geometric constraints on the accessible phase space of the composite system but does not actively participate in pair-wise interactions.
This assumption isolates the solvent-solvent coupling as the only source of system-environment interaction. Furthermore, we do not incorporate any internal degrees of freedom for the solvent molecules. That is, they are treated as structureless particles that interact solely via a pairwise Lennard-Jones potential as shown in Fig.~\ref{solution}.
This simplification focuses the analysis on translational degrees of freedom and excludes intramolecular complexity from the statistical treatment. It is important, however, to emphasize that neglecting solute–outer-shell interactions is introduced here solely to reduce the notational complexity of the derivations and does not limit the generality of the formal results. 
Rather, the framework developed in this work remains valid and can be systematically extended to include solute–outer-shell interactions. 
A motivated reader may use the same structure to straightforwardly generalize the equations accordingly.
In future work, we will explicitly demonstrate how the inclusion of such interactions within the present framework enables the computation of the excess chemical potential for solute particles exhibiting non-hard-core interactions.\\
The microscopic state of each region is described by phase-space variables: $x_\mathcal{S} = (\mathbf{P}_{\mathcal{S}}, \mathbf{Q}_{\mathcal{S}})$ for the system and $x_\mathcal{E} = (\mathbf{P}_{\mathcal{E}}, \mathbf{Q}_{\mathcal{E}})$ for the environment. 
A central question in statistical mechanics is how macroscopic behavior emerges from a microscopic observable $\mathcal{A}(x_\mathcal{S})$ that depends solely on the degrees of freedom of $\mathcal{S}$. 
In this context, the role of statistical mechanics is to relate such observables to thermodynamic quantities through a probability distribution $P(x_\mathcal{S})$, which assigns statistical weights to different configurations of the system leading to the ensemble average $\langle\mathcal{A}\rangle$ of observable $\mathcal{A}$
\begin{align}
\langle\mathcal{A}\rangle = \int \mathcal{A}(x_\mathcal{S})\, P(x_\mathcal{S})\, \mathrm{d} x_\mathcal{S}
\end{align}
To evaluate this expectation, we must determine the marginal distribution $P(x_\mathcal{S})$, obtained by integrating out the environmental degrees of freedom
\begin{equation}
P(x_\mathcal{S}) = \int P(x_\mathcal{S}, x_\mathcal{E})\, \mathrm{d}x_\mathcal{E}.
\end{equation}
This perspective aligns with the logic behind the Hamiltonian of mean force (HMF) model, which captures system behavior by incorporating the environmental degrees of freedom~\cite{feynman2000theory,kirkwood1935statistical,roux1999implicit,talkner2016open,talkner2020colloquium,jarzynski2017stochastic,miller2018energy,seifert2016first,miller2017entropy,talkner2016open,talkner2020colloquium,anto2023effective}.
Inspired by this approach, we develop a framework in which the system-environment interaction $V_{\mathcal{S}\mathcal{E}}(x_\mathcal{S}, x_\mathcal{E})$ is treated as a random variable.
In the following, we systematically show how the interaction-energy distribution $P(V_{\mathcal{S}\mathcal{E}})$ can be directly connected to the marginal distribution $P(x_\mathcal{S})$. 
To begin, it is essential to recognize that the composite system $\mathcal{S}+\mathcal{E}$ is assumed to be in equilibrium in the thermodynamic limit.
As a consequence, the joint distribution $P(x_\mathcal{S}, x_\mathcal{E})$ becomes effectively independent of the ensemble used to describe the full system of particles \cite{kardar2007statistical,jarzynski2017stochastic}.
Since all ensembles are equivalent in the thermodynamic limit, the choice is guided purely by mathematical convenience. 
Hence, in this work, the composite system can be treated using either a microcanonical ensemble --- where total energy $E$, particle number $N$, and volume $V$ are fixed --- or a canonical ensemble, in which the composite system is embedded in a thermal reservoir at temperature $T$. 
This equivalence further implies that the marginal distribution $P(x_\mathcal{S})$ remains invariant under the choice of ensemble \cite{jarzynski2017stochastic}.
Regarding the constraints on the system, and following the earlier discussion, we assume that the system can exchange both energy and particles with its environment.
In particular, no restriction is imposed on the number or configuration of solvent particles occupying the inner shell. 
When working in the microcanonical picture, the total energy is fixed, and only those configurations that satisfy the energy constraint contribute to the ensemble. 
This requirement is captured by the condition \cite{kardar2007statistical}
\begin{align}  
\delta \mathcal{C}(x_\mathcal{E}, x_\mathcal{S}; E) &\nonumber:= \delta(E - \mathcal{H}_{\mathcal{E}}(x_\mathcal{E}) - \mathcal{H}_{\mathcal{S}}(x_\mathcal{S}) \\[6pt]
&- V_{\mathcal{S}\mathcal{E}}(x_\mathcal{S}, x_\mathcal{E})),
\label{eq:energy_constraint}
\end{align}
 here $\mathcal{H}_{\mathcal{E}}(x_\mathcal{E})$ and $\mathcal{H}_{\mathcal{S}}(x_\mathcal{S})$ are the Hamiltonians of $\mathcal{E}$ and $\mathcal{S}$, respectively. This condition ensures that each configuration of $\mathcal{S} + \mathcal{E}$ lies on the constant-energy hypersurface since the $\delta(\cdot)$ function ensures that only configurations compliant with the total energy $E$ are included. 
 To classify microstates further according to their coupling strength, measured by the interaction energy, we introduce the following composite constraint
\begin{align}
\Delta(x_\mathcal{E}, x_\mathcal{S}; E, V_{\mathcal{S}\mathcal{E}})&\nonumber := \delta \mathcal{C}(x_\mathcal{E}, x_\mathcal{S}; E) \\[6pt]
&\times \delta(V_{\mathcal{S}\mathcal{E}}(x_\mathcal{S}, x_\mathcal{E}) - V_{\mathcal{S}\mathcal{E}}),
\label{eq:delta_combined}
\end{align}
which filters configurations based on both total energy and the value of the interaction term. 
To evaluate statistical averages over all microscopic realizations of the composite system $\mathcal{S} + \mathcal{E}$, we rely on a trace operator that accounts for both particle partitioning and phase-space integration.
The operator $\mathcal{L}^N\{\bullet\}$ provides a compact way to handle this, and is defined as
\begin{align}
\mathcal{L}^{N} \{\bullet\} := \sum_{N_\mathcal{S}=0}^{N} \sum_{(I,J)_{\emptyset}} 
\prod_{m \in I} \int_{\mathcal{S}} \mathrm{d}x_m 
\prod_{n \in J} \int_{\mathcal{E}} \mathrm{d}x_n 
\{\bullet\},
\label{eq:notation_trace}
\end{align}
where
\begin{equation}
\int_{\mathcal{S}} \mathrm{d}x_m  = \int_{-\infty}^{\infty}\int_{V_\mathcal{S}}\mathrm{d}q_m \mathrm{d}p_m , \quad \int_{\mathcal{E}} \mathrm{d}x_n  = \int_{-\infty}^{\infty}\int_{V_\mathcal{E}} \mathrm{d}q_n \mathrm{d}p_n.
\end{equation}
Here, the placeholder $\bullet$ represents any observable or constraint defined on the full phase space. 
The outer summation runs over all possible particle counts in the system, from $0$ to $N$, while the inner sum $\sum_{(I,J)_\emptyset}$ enumerates all disjoint partitions of the phase-space variables of all $N$ particles into two subsets: $I$ assigned to $\mathcal{S}$ and $J$ to $\mathcal{E}$. 
Each such partition satisfies
\begin{equation}
(I,J)_{\emptyset} := \left\{ (I, J) \mid I \in \mathbf{S}_{\binom{N}{N_\mathcal{S}}}, \; J \in \mathbf{E}_{\binom{N}{N - N_\mathcal{S}}}, \; I \cap J = \emptyset \right\},
\end{equation}
ensuring mutual exclusivity between system and environment assignments.
The sets $\mathbf{S}$ and $\mathbf{E}$ enumerate all possible combinations of particle variables for each region.
Since the trace operator $\mathcal{L}^{N} \{\bullet\}$ systematically generates all admissible configurations based on the topology of the system and the environment, it inherently encodes which particles belong to $\mathcal{S}$ and which to $\mathcal{E}$ in each configuration.
By this construction, every generated microstate carries an explicit system-environment assignment.
Suppose we consider a composite system without explicitly distinguishing between the system and the environment. 
Within this unified representation, we can construct arbitrary configurations without imposing predefined roles on particles. 
Then, in evaluating any observable that depends on the particle assignment, we introduce a detector function $\mathcal{I}{\{\dots\}}$ --- a purely abstract construct --- that identifies whether a given particle resides in $\mathcal{S}$ or $\mathcal{E}$. 
The detector function can be defined formally as
\begin{equation}
\mathcal{I}{\{\mathcal{S}\}}: \Gamma_N \rightarrow L_{\mathcal{S}}, \qquad
\mathcal{I}{\{\mathcal{E}\}}: \Gamma_N \rightarrow L_{\mathcal{E}},
\end{equation}
Here, $L_{\mathcal{S}}$ and $L_{\mathcal{E}}$ are the sets of particle labels detected within regions $\mathcal{S}$ and $\mathcal{E}$, respectively, while $\Gamma_N$ denotes the full $N$-particle phase space. 
Then, for any configuration $x = (\mathbf{P, \mathbf{Q}}) \in \Gamma_N$, the detector functions $\mathcal{I}\{\mathcal{S}\}$ and $\mathcal{I}\{\mathcal{E}\}$ return the label subsets corresponding to particles residing in $\mathcal{S}$ and $\mathcal{E}$, respectively. 
In this way, the system-environment partition is detected directly from the configuration itself, rather than generated externally. 
In parallel, we define the following mappings
\begin{equation}
\mathcal{I}\{\mathcal{S}X\}: \Gamma_N \rightarrow X_\mathcal{S}, \qquad
\mathcal{I}\{\mathcal{E}X\}: \Gamma_N \rightarrow X_\mathcal{E},
\end{equation}
where $X_\mathcal{S}$ and $X_\mathcal{E}$ denote the sub-configurations (phase-space components) of the particles that belong to the system and environment, respectively.
These mappings extract the relevant phase-space degrees of freedom directly from the full configuration $x$.
Using this construction, we introduce a modified trace operator that incorporates detector-based classification
\begin{align}
\mathcal{IL}^{N} \{\bullet\} 
&\nonumber:= \int_{\Gamma_N} \mathrm{d}x \, \mathcal{I}\{\mathcal{S}X,\mathcal{E}X\}(x) \, \{\bullet\}\\&:=\int_{\Gamma_N} \mathrm{d}\mathbf{P}\mathrm{d}\mathbf{Q} \, \mathcal{I}\{\mathcal{S}X,\mathcal{E}X\}(\mathbf{P, \mathbf{Q}}) \, \{\bullet\}\label{IL}
\end{align}
where the combined detector $\mathcal{I}\{\mathcal{S}X,\mathcal{E}X\}$ acts on the full configuration $x$ to select and assign the relevant degrees of freedom to system and environment subspaces before evaluating the observable $\{\bullet\}$. 
This formulation enables us to treat the composite system without imposing a fixed boundary between the system and environment.
Instead, the classification of particles is deferred until the point of evaluation, where it is determined by the detector function. 
It is crucial to emphasize that the choice of representation does not affect the final result. 
Regardless of whether we analyze a given observable using explicit system-environment labeling or through the detector-based approach, the outcome remains the same. 
That is,
\begin{equation}
\mathcal{IL}^{N} \{\bullet\} = \mathcal{L}^{N} \{\bullet\}.
\end{equation}
It means that the two frameworks --- despite their conceptual differences --- yield identical statistical predictions when applied to observables defined over the full configuration space.
To filter configurations in which particle $k$ occupies a specific position $q_\odot$ in the system, we impose the constraint via a delta function
\begin{align}
\mathcal{IL}^{N} \left\{ \delta(q_k - q_\odot)\, \bullet \right\} \nonumber&:=  \mathcal{IL}^{\circledcirc}_{q_k} \left\{\delta_{q_\odot q_k} \otimes\bullet\right\} \\&:= \delta_{q_\odot q_k} \otimes\mathcal{IL}^{\circledcirc}_{q_k} \{\bullet\}.
\label{eq:delta_prop}
\end{align}
The essential action of $\delta_{q_\odot q_k} \otimes$ can be understood in simple terms: when applied to a function, it substitutes every instance of $q_k$ with the fixed position $q_\odot$. For example, if the operator acts on a function of all particle coordinates such as $f(q_1, \dots, q_k, \dots, q_N)$, then
\begin{equation}
\delta_{q_\odot q_k} \otimes f(q_1, \dots, q_k, \dots, q_N) = f(q_1, \dots, q_\odot, \dots, q_N).
\end{equation}
Physically, this corresponds to evaluating the observable under the condition that particle $k$ occupies a fixed spatial location.
Mathematically, the operator is distributive over both addition and multiplication. That is, for two arbitrary functions $f$ and $g$, we have 
\begin{align}
\delta_{q_\odot q_k} \otimes (f + g) &= (\delta_{q_\odot q_k} \otimes f) + (\delta_{q_\odot q_k} \otimes g), \\[4pt]
\delta_{q_\odot q_k} \otimes (f \cdot g) &= (\delta_{q_\odot q_k} \otimes f) \cdot (\delta_{q_\odot q_k} \otimes g).
\end{align}
These properties ensure that $\delta_{q_\odot q_k} \otimes$ behaves as a linear substitution operator across composite expressions, allowing it to be applied consistently in more complex integrals and product forms. 
It is also important to note that the operator $\mathcal{IL}^{\circledcirc}_{q_k}$ refers to a modified version of $\mathcal{IL}^{N}$ in which integration over the coordinate $q_k$ is intentionally omitted. 
The use of the superscript $\circledcirc$ signals that all degrees of freedom are included in the trace except for $q_k$. 
Using the established notations, we can express the total phase-space volume accessible to the combined system at fixed energy $E$ as
\begin{equation}
\Omega = \mathcal{IL}^{N} \{ \delta \mathcal{C}(x_\mathcal{E}, x_\mathcal{S}; E) \}.
\label{eq:omega}
\end{equation}
Here, $\Omega$ captures the total number of microstates consistent with the energy constraint across all possible system-environment partitions. 
To help the reader to better grasp the operational meaning behind the detector-based operator $\mathcal{IL}^{N}$, we present an explicit example by applying it to the energy constraint
\begin{align}
\mathcal{IL}^{N} \delta \mathcal{C}(x_\mathcal{E}, x_\mathcal{S}; E)  
&= \int_{\Gamma_N} \mathrm{d}x \, \delta\Bigl( E - \mathcal{H}_{\mathcal{E}}(\mathcal{I}\{\mathcal{E}X\}(x)) \nonumber\\[4pt]
&\quad - \mathcal{H}_{\mathcal{S}}(\mathcal{I}\{\mathcal{S}X\}(x)) \nonumber\\[4pt]
&\quad - V_{\mathcal{S}\mathcal{E}}(\mathcal{I}\{\mathcal{S}X\}(x), \mathcal{I}\{\mathcal{E}X\}(x)) \Bigr),
\end{align}
where the detector mappings $\mathcal{I}\{\mathcal{S}X\}(x)$ and $\mathcal{I}\{\mathcal{E}X\}(x)$ extract the appropriate sub-configurations from the full phase-space point $x \in \Gamma_N$. 
Although we often use the compact notation $\mathcal{IL}^{N}\{\bullet\}$ throughout this work, it is essential to interpret such expressions in the operational sense illustrated above. When the system configuration $x_\mathcal{S}$ is specified, its probability is determined by how many environmental microstates are compatible with it
\begin{equation}
P(x_\mathcal{S}) = \frac{\int \mathrm{d}{x_\mathcal{E}} \,\delta \mathcal{C}(x_\mathcal{E}, x_\mathcal{S}; E) }{ \Omega }.
\label{eq:probability}
\end{equation}
The interaction energy $V_{\mathcal{S}\mathcal{E}}$ generally fluctuates from one configuration to another. To account for this explicitly, we treat it as a random variable and decompose the marginal probability accordingly
\begin{equation}
P(x_\mathcal{S}) = \int dV_{\mathcal{S}\mathcal{E}}\, P(x_\mathcal{S} \mid V_{\mathcal{S}\mathcal{E}})\, P(V_{\mathcal{S}\mathcal{E}}).
\label{eq:total_prob}
\end{equation}
The conditional term $P(x_\mathcal{S} \mid V_{\mathcal{S}\mathcal{E}})$ reflects the likelihood of a system configuration given that the interaction energy has a specified value. It is computed by tracing over environmental degrees of freedom under both energy and interaction constraints:
\begin{equation}
P(x_\mathcal{S} \mid V_{\mathcal{S}\mathcal{E}}) = \frac{\int \mathrm{d}{x_\mathcal{E}} \, \Delta(x_\mathcal{E}, x_\mathcal{S}; E, V_{\mathcal{S}\mathcal{E}}) }{ \mathcal{IL}^{N} \left\{ \Delta(x_\mathcal{E}, x_\mathcal{S}; E, V_{\mathcal{S}\mathcal{E}}) \right\} },
\label{eq:cond_prob}
\end{equation}
where $\Delta$ enforces both constraints simultaneously [see Eq.~\ref{eq:delta_combined}]. The probability distribution for the interaction energy itself is given as
\begin{equation}
P(V_{\mathcal{S}\mathcal{E}}) = \frac{ \mathcal{IL}^{N} \{ \Delta(x_\mathcal{E}, x_\mathcal{S}; E, V_{\mathcal{S}\mathcal{E}}) \} }{ \Omega }.
\label{eq:interaction_prob}
\end{equation}
Together, these components define a consistent framework for analyzing how system statistics are shaped by the underlying interaction. 
In what follows, we build on this foundation to quantify the role of $V_{\mathcal{S}\mathcal{E}}$ in controlling fluctuations and determining the effective strength of coupling between $\mathcal{S}$ and $\mathcal{E}$.
\section{Reference Particle Approach}
\label{rpa}
We now introduce a reference-particle approach to compute statistical quantities related to system-environment interactions.
To illustrate the idea, suppose we are interested in the average of a function $\mathcal{A}_{\mathcal{S}\mathcal{E}}(x_\mathcal{S}, x_\mathcal{E})$ that depends explicitly on the positions of particles in both the system and the environment. For instance, the first and second moments of the interaction energy can be viewed as specific examples of the general structure captured by $\langle \mathcal{A}_{\mathcal{S}\mathcal{E}} \rangle$. The framework introduced earlier enables us to express such averages in the form
\begin{equation}
\langle \mathcal{A}_{\mathcal{S}\mathcal{E}} \rangle = \frac{\mathcal{IL}^{N} \{ \delta \mathcal{C}(x_\mathcal{E}, x_\mathcal{S}; E) \mathcal{A}_{\mathcal{S}\mathcal{E}}(x_\mathcal{S}, x_\mathcal{E})\}}{\Omega}.
\label{eq:interaction_expectation_initial}
\end{equation}
 In practice, computing the expectation value in Eq.~(\ref{eq:interaction_expectation_initial}) is rarely straightforward.
 The challenge lies in the fact that $\mathcal{A}_{\mathcal{S}\mathcal{E}}$ depends on the complete microscopic configuration of both the system and the environment. 
 Consequently, the associated integral must account for all degrees of freedom of all particles --- an operation that becomes quickly intractable as the number of particles grows. 
 The reference-particle approach offers a practical way to circumvent this complexity. 
 By focusing on a single, tagged particle in the system, we can decompose $\mathcal{A}_{\mathcal{S}\mathcal{E}}$ into contributions that are far more tractable. 
 We assume that $\mathcal{A}_{\mathcal{S}\mathcal{E}}$ can be written as a sum over reference particles
\begin{equation}
\mathcal{A}_{\mathcal{S}\mathcal{E}}(x_\mathcal{S}, x_\mathcal{E}) = \sum_{k \in \mathcal{I}{\{\mathcal{S}\}}(x)} \mathcal{A'}_{\mathcal{S}\mathcal{E}}(q_k, \circledcirc).
\label{decom1}
\end{equation}
Here, $\mathcal{A'}_{\mathcal{S}\mathcal{E}}(q_k, \circledcirc)$ captures the contribution associated with particle $k$ located at $q_k$ within the system, while $\circledcirc$ denotes the remaining degrees of freedom of all particles in both the system and the environment. 
This decomposition proves particularly useful when dealing with observables that describe system-environment interactions or similar coupled quantities. 
For instance, if $\mathcal{A}_{\mathcal{S}\mathcal{E}}$ quantifies a system-environment interaction, then $\mathcal{A'}_{\mathcal{S}\mathcal{E}}$ characterizes how a single particle in the system interacts with the surrounding environment. 
Analyzing the environment's response to an individual particle is often more tractable and conceptually transparent. 
Moreover, this formulation naturally leads to the definition of single-particle densities and multi-body correlation functions—statistical objects that provide quantitative insight into how local interactions shape global behavior \cite{hansen2013theory}.
It is important to note that the functional form of $\mathcal{A'}_{\mathcal{S}\mathcal{E}}(q_k, \circledcirc)$ is not uniquely determined and generally depends on the specific structure of $\mathcal{A}_{\mathcal{S}\mathcal{E}}$, a point we will discuss in detail later.
For the current discussion, however, it is not necessary to specify an explicit form.
What matters is the assumption that such a decomposition exists and that $\mathcal{A'}_{\mathcal{S}\mathcal{E}}(q_k, \circledcirc)$ depends on the complete microscopic configuration of the composite system. 
More precisely, the pair $(q_k, \circledcirc)$ is understood to include all degrees of freedom in both the system and the environment. 
Inserting Eq.~(\ref{decom1}) into Eq.~(\ref{eq:interaction_expectation_initial}), we can write
\begin{align}
\langle \mathcal{A}_{\mathcal{S}\mathcal{E}} \rangle =\frac{1}{\Omega} \mathcal{IL}^{N}\left\{\delta \mathcal{C}(x_\mathcal{E}, x_\mathcal{S}; E)\sum_{k \in \mathcal{I}{\{\mathcal{S}\}}(x)} \mathcal{A'}_{\mathcal{S}\mathcal{E}}(q_k, \circledcirc)\right\} .
\label{genral ave}
\end{align}
As already implied by the structure of $\mathcal{IL}^{N}$, the collection $\mathcal{I}{\{\mathcal{S}\}}(x)$ should not be regarded as a fixed set of particles. 
Instead, it varies from configuration to configuration, both in the number of particles assigned to the system and in their identities (i.e., labels). 
As a result, the notation $\sum_{k \in \mathcal{I}{\{\mathcal{S}\}}(x)} $ can only be regarded as a symbolic, qualitative prescription rather than a mathematically well-defined operation.
To address this issue, the idea is to  reformulate the summation over system particles in terms of operation over all particles in the composite configuration. 
Using the properties of the Dirac delta function, we can rewrite the summation over system-detected particles in terms of the full configuration as
\begin{align}
\sum_{k \in \mathcal{I}{\{\mathcal{S}\}(x)}}  \longrightarrow  \sum_{k=1}^N \int_{V_\mathcal{S}} \mathrm{d} q_\odot \, \delta(q_\odot - q_k).
\label{transformation}
\end{align}
In this representation, the delta function acts as a filter: for any given configuration, it identifies the particles whose positions reside within the system volume. 
In what follows, we will repeatedly make use of this transformation. 
It is important to emphasize that this substitution is not only powerful but also highly extensible. 
As an example, we can transform $\sum_{k \in \mathcal{I}{\{\mathcal{S}\}}}  \sum_{j \in \mathcal{I}{\{\mathcal{E}\}}} $ to
\begin{align}
\sum_{k=1}^N \sum_{\substack{j=1\\ j\neq k}}^{N}\int_{V_\mathcal{S}} \mathrm{d} q_\odot \int_{V_\mathcal{E}} \mathrm{d} q \, \delta(q_\odot - q_k)\delta(q - q_j).
\label{powertrans}
\end{align}
Applying the transformation in Eq.~(\ref{transformation}) to the decomposition in Eq.~(\ref{decom1}), we obtain
\begin{equation}
\mathcal{A}_{\mathcal{S}\mathcal{E}}(x_\mathcal{S}, x_\mathcal{E}) = \sum_{k=1}^N \int_{V_\mathcal{S}} \mathrm{d} q_\odot \, \delta(q_\odot - q_k) \, \mathcal{A'}_{\mathcal{S}\mathcal{E}}(q_k, \circledcirc).
\end{equation}
Substituting this result into Eq.~(\ref{genral ave}) yields the following expression for the expectation value
\begin{align}
\langle \mathcal{A}_{\mathcal{S}\mathcal{E}} \rangle 
\nonumber&= \frac{1}{\Omega} \int_{V_\mathcal{S}} \mathrm{d} q_\odot \sum_{k=1}^N \mathcal{IL}^{N} \{ \delta \mathcal{C}(x_\mathcal{E}, x_\mathcal{S}; E) \, \delta(q_\odot - q_k) \\&\times\, \mathcal{A'}_{\mathcal{S}\mathcal{E}}(q_k, \circledcirc) \}.
\label{refexp}
\end{align}
As shown in detail in the Appendix~\ref{Arpa}, Eq.~(\ref{refexp}) can be written in the following compact form
\begin{equation}
\langle \mathcal{A}_{\mathcal{S}\mathcal{E}} \rangle = \int_{V_\mathcal{S}} \mathrm{d}q_\odot \, \rho(q_\odot) \, \langle \mathcal{A'}_{\mathcal{S}\mathcal{E}}(q_\odot, \circledcirc) \rangle_{\circledcirc|q_\odot}.
\end{equation}
Here, $\rho(q_\odot)$ denotes the generic single-particle probability density: the probability of finding a particle at position $q_\odot$, regardless of its identity or label.
The second factor, $\langle \mathcal{A'}_{\mathcal{S}\mathcal{E}}(q_\odot, \circledcirc) \rangle_{\circledcirc|q_\odot}$, represents the conditional expectation of the localized contribution $\mathcal{A'}_{\mathcal{S}\mathcal{E}}$, given that a particle is fixed at $q_\odot$. 
This quantity is computed by averaging over all remaining degrees of freedom, while holding a particle always fixed at $q_\odot$. 
In the case where particles are assumed to be identical, one can show that the conditional expectation simplifies as
\begin{equation}
\langle \mathcal{A'}_{\mathcal{S}\mathcal{E}}(q_\odot, \circledcirc) \rangle_{\circledcirc|q_\odot} 
= \delta_{q_\odot q_k} \otimes \langle \mathcal{A'}_{\mathcal{S}\mathcal{E}}(q_k,\circledcirc) \rangle_{\circledcirc | q_k},
\label{eq:final_result1}
\end{equation}
where we introduce $\langle \mathcal{A'}_{\mathcal{S}\mathcal{E}}(q_k,\circledcirc) \rangle_{\circledcirc | q_k}$ as the specific single-particle conditional expectation. 
Thus, with Eq.~(\ref{eq:final_result1}), the expectation of the interaction-like observable can be expressed as
\begin{align}
\langle \mathcal{A}_{\mathcal{S}\mathcal{E}} \rangle 
= \int_{V_\mathcal{S}} \mathrm{d}q_\odot \, \rho(q_\odot) \, \delta_{q_\odot{q}_{k}} \otimes \langle\mathcal{A'}_{\mathcal{S}\mathcal{E}}(q_k,\circledcirc) \rangle_{\circledcirc|q_k}.
\label{central}
\end{align}
It is important to note that, since the particles are identical, the index $k$ in the above expression is arbitrary. 
That is, $k$ can represent any particle located at position $q_\odot$, as all particles contribute equivalently since a permutation of particles of equal kind does not alter the interaction. 
It can be shown (see Appendix~\ref{Arpa}) that the specific single-particle conditional expectation can be written as
\begin{equation}
\langle \mathcal{A'}_{\mathcal{S}\mathcal{E}}(q_k,\circledcirc) \rangle_{\circledcirc \mid q_k} 
= \mathcal{IL}_{q_k}^{\circledcirc} \left\{ P(\circledcirc \mid q_k) \, \mathcal{A'}_{\mathcal{S}\mathcal{E}}(q_k,\circledcirc) \right\}.
\label{eq:single_particle_cond_exp}
\end{equation}
To clarify how the single-particle conditional expectation is computed, suppose that the observable $\mathcal{A'}_{\mathcal{S}\mathcal{E}}(q_k,\circledcirc)$ can be expressed as a function $\mathcal{B}(q_k, \mathbf{Q})$ (a  common scenario for all interaction-like quantities), where $\mathbf{Q}$ denotes the positions of all particles except for particle $k$. 
In this case, applying Bayes’ rule and decomposing the conditional probability into more tractable terms simplifies the evaluation. 
In particular, the conditional probability $P(\circledcirc \mid q_k)$—which captures the distribution over all degrees of freedom excluding $q_k$, conditioned on the fact that $q_k$ is fixed, can be rewritten as
\begin{align}
P(\circledcirc | q_k) &= \frac{P(\circledcirc, q_k)}{P(q_k)} = \frac{P(\circledcirc \backslash \mathbf{Q}, \mathbf{Q}, q_k)}{P(q_k)} \nonumber \\
&= P(\circledcirc \backslash \mathbf{Q} | \mathbf{Q}, q_k) P(\mathbf{Q} | q_k). \label{eq:conditional_prob_hierarchy}
\end{align}
Here, $\circledcirc \backslash \mathbf{Q}$ represents the whole coordinate space excluding the position of all particles.
With this, the single-particle conditional expectation can be expanded as
\begin{align}
\langle \mathcal{A'}_{\mathcal{S}\mathcal{E}}(q_k,\circledcirc) \rangle_{\circledcirc \mid q_k} &\nonumber =
\mathcal{IL}_{q_k}^{\circledcirc \backslash \mathbf{Q}}\{P(\circledcirc \backslash \mathbf{Q} | \mathbf{Q}, q_k)\} \\&\times\mathcal{IL}_{q_k}^{\mathbf{Q}}\{P(\mathbf{Q} | q_k) \mathcal{B}(q_k, \mathbf{Q})\} 
\label{eq:expanded_cond_expectation}
\end{align}
By decomposing the integral over $\circledcirc$ into two subsets, we can isolate the relevant variables $\mathbf{Q}$ and integrate out the rest. 
Thus, we can write $\circledcirc = \mathbf{Q} \cup (\circledcirc \setminus \mathbf{Q})$, and see that
\begin{equation}
\mathcal{IL}_{q_k}^{\circledcirc \setminus \mathbf{Q}} \left\{ P(\circledcirc \setminus \mathbf{Q} \mid \mathbf{Q}, q_k) \right\} = 1.
\end{equation}
Substituting this result into the definition of the conditional expectation, we obtain
\begin{align}
\mathcal{IL}_{q_k}^{\circledcirc} \left\{ P(\circledcirc \mid q_k) \, \mathcal{B}(\mathbf{Q}) \right\}
= \int \mathrm{d}\mathbf{Q} \, P(\mathbf{Q} \mid q_k) \, \mathcal{B}(\mathbf{Q}),
\label{eq:reduced_cond_expectation}
\end{align}
which expresses the conditional expectation in terms of a reduced subset of degrees of freedom. This hierarchical decomposition of probabilities also enables us to isolate smaller subsets of variables and marginalize over the remaining ones. 
In general, if we are interested in computing an expectation of the form 
$\mathcal{IL}_{q_k}^{\circledcirc} \left\{ P(\circledcirc \mid q_k) \, \mathcal{B}(\mathbf{X'}) \right\}$, 
where $\mathbf{X'} \subset \circledcirc$, we may proceed analogously and write
\begin{align}
\mathcal{IL}_{q_k}^{\circledcirc} \left\{ P(\circledcirc \mid q_k) \, \mathcal{B}(\mathbf{X'}) \right\} 
= \int \mathrm{d}\mathbf{X'} \, P(\mathbf{X'} \mid q_k) \, \mathcal{B}(\mathbf{X'}).
\label{eq:reduced_cond_expectation}
\end{align}
One important point that to emphasize is that, throughout this work, we have assumed that $\mathcal{A}_{\mathcal{S}\mathcal{E}}$ is a function of the full set of degrees of freedom in both $\mathcal{S}$ and $\mathcal{E}$.
This assumption is completely general and provides a unified representation that can be adapted to specific cases when needed. For example, in most practical scenarios, the interaction energy $V_{\mathcal{S}\mathcal{E}}$ typically depends only on the particle positions, not on their momenta.
That is, it takes the form $V_{\mathcal{S}\mathcal{E}}(\mathbf{Q}_\mathcal{S}, \mathbf{Q}_\mathcal{E})$ rather than depending on the full phase-space coordinates $(x_\mathcal{S}, x_\mathcal{E})$. 
However, for the sake of generality and to ensure that our derivations hold under the strongest assumptions, we have adopted the more inclusive notation $V_{\mathcal{S}\mathcal{E}}(x_\mathcal{S}, x_\mathcal{E})$ in all intermediate steps. 
When necessary, this general representation can be readily specialized to match the functional form relevant to the physical system under consideration. This distinction becomes particularly important in the subsequent analysis of expectation values and fluctuations of the interaction energy, where the physical dependence on particle positions becomes more explicit and interpretable.

\section{Derivation of the first moment of Interaction}\label{DFM}
\begin{figure}[htbp]
    \centering
    \includegraphics[width=\columnwidth]{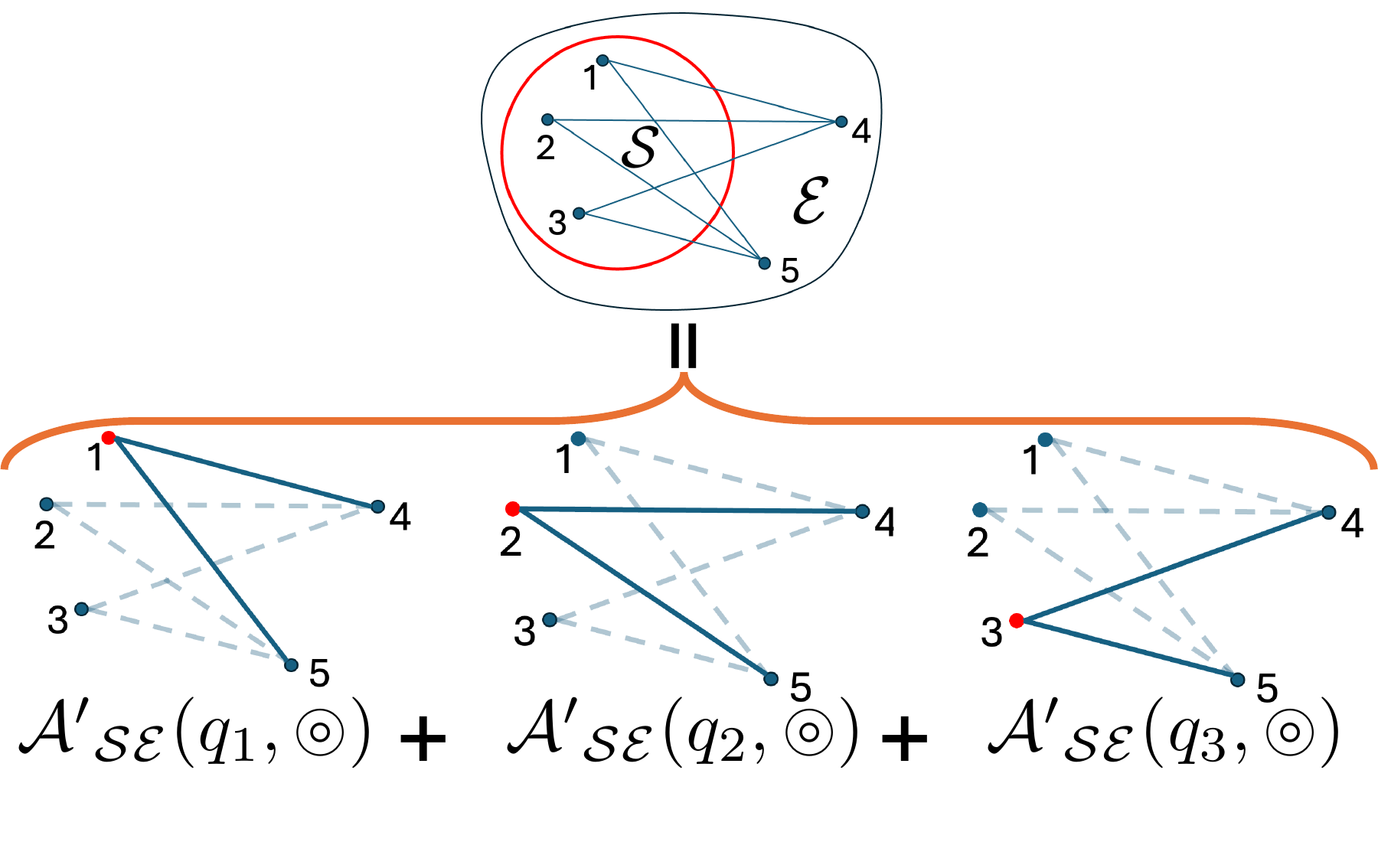}
    \caption{Decomposition of the total interaction potential $V_{\mathcal{S}\mathcal{E}}$ into contributions from each reference particle (cf. Eq.~(\ref{eq:four})).}
    \label{fig:total interaction potential}
\end{figure}

In the following, we turn to the explicit computation of the average interaction energy $\langle V_{\mathcal{S}\mathcal{E}} \rangle$. 
As discussed above, we begin with the most general representation, treating $V_{\mathcal{S}\mathcal{E}}$ as a function of the full set of degrees of freedom in both $\mathcal{S}$ and $\mathcal{E}$. 
That is, we initially assume 
$V_{\mathcal{S}\mathcal{E}} = V_{\mathcal{S}\mathcal{E}}(x_\mathcal{S}, x_\mathcal{E})$. 
Once the general expressions are derived, we specialize to the physically relevant case where the interaction energy depends only on the positions of the particles, i.e., 
$V_{\mathcal{S}\mathcal{E}} = V_{\mathcal{S}\mathcal{E}}(\mathbf{Q}_\mathcal{S}, \mathbf{Q}_\mathcal{E})$. 
With the blueprint given in Eq.~(\ref{eq:interaction_expectation_initial}), the average interaction energy can formally be written as
\begin{equation}
\langle V_{\mathcal{S}\mathcal{E}} \rangle = \frac{\mathcal{IL}^{N} \left\{ \delta \mathcal{C}(x_\mathcal{E}, x_\mathcal{S}; E) \,V_{\mathcal{S}\mathcal{E}}(x_\mathcal{S}, x_\mathcal{E}) \right\}}{\Omega}.
\label{eq:one}
\end{equation}
In many relevant physical scenarios, such as non-bonding interactions (e.g., van der Waals or Coulomb interactions), the total interaction energy can be expressed as a sum over pairwise interactions between particles located in the system and those in the environment. 
That is,
\begin{equation}
V_{\mathcal{S}\mathcal{E}}(x_{\mathcal{E}}, x_{\mathcal{S}}) = \sum_{k \in \mathcal{I}{\{\mathcal{S}\}(x)}}   \sum_{j \in \mathcal{I}{\{\mathcal{E}\}(x)}}  u(\| q_k - q_j \|_2).
\label{eq:two}
\end{equation}
where $u(q_k, q_j)$ denotes the pair potential between particle $k$ in the system and particle $j$ in the environment. Based on the algebraic structure introduced in Eq.~(\ref{eq:two}), it is straightforward to express the interaction energy in terms of contributions from individual system particles. 
Specifically, we write
\begin{equation}
V_{\mathcal{S}\mathcal{E}}(x_\mathcal{S}, x_\mathcal{E}) = \sum_{k \in \mathcal{I}\{\mathcal{S}\}(x)} \mathcal{A'}_{\mathcal{S}\mathcal{E}}(q_k, \circledcirc),
\label{eq:three}
\end{equation}
where
\begin{equation}
\mathcal{A'}_{\mathcal{S}\mathcal{E}}(q_k, \circledcirc) =  \sum_{j \in \mathcal{I}{\{\mathcal{E}\}(x)}} u(\| q_k - q_j \|_2).
\label{eq:four}
\end{equation}
Each term $\mathcal{A'}_{\mathcal{S}\mathcal{E}}(q_k, \circledcirc)$ describes the contribution to the total interaction energy associated with a single reference particle $k$ in the system.
To make the algebraic structure more transparent, we introduce a schematic representation of our reference particle approach in Fig.~\ref{fig:total interaction potential}). 

In this example, we consider a small composite system $\mathcal{S} + \mathcal{E}$ consisting of a total of five particles. 
Naturally, the underlying decomposition is valid for any configuration and an arbitrary number of particles. 
In this setting, the total interaction energy $V_{\mathcal{S}\mathcal{E}}$ is expressed as the sum of three contributions of the form $\mathcal{A'}_{\mathcal{S}\mathcal{E}}(q_k, \circledcirc)$ that capture the interaction of particle $k$ with all particles in the environment $\mathcal{E}$. 
With all the necessary ingredients now in place, we are ready to express the average interaction energy in its final, operational form. 
Following Eq.~(\ref{central}), we write
\begin{equation}
\langle V_{\mathcal{S}\mathcal{E}} \rangle = \int_{V_\mathcal{S}} \mathrm{d}q_\odot \, \rho(q_\odot) \, \delta_{q_\odot q_k} \otimes \langle \mathcal{A'}_{\mathcal{S}\mathcal{E}}(q_k, \circledcirc) \rangle_{\circledcirc | q_k}.
\label{V_expect}
\end{equation}
The interaction contribution $\mathcal{A'}_{\mathcal{S}\mathcal{E}}(q_k, \circledcirc)$, as introduced in Eq.~(\ref{eq:four}), can be transformed with the transformation identity derived in Eq.~(\ref{powertrans}) to yield
\begin{align}
\mathcal{A'}_{\mathcal{S}\mathcal{E}}(q_k, \circledcirc)
= \sum_{\substack{j=1 \\ j \neq k}}^{N} \int_{V_\mathcal{E}} \mathrm{d}q \, \delta(q - q_j) \, u\big( \| q - q_k \|_2 \big),
\label{A'inter}
\end{align}
To proceed, we focus on the conditional expectation $\langle \mathcal{A'}_{\mathcal{S}\mathcal{E}}(q_k, \circledcirc) \rangle_{\circledcirc | q_k}$.
Leveraging Eq.~(\ref{eq:single_particle_cond_exp}) along with the transformed representation introduced in Eq.~(\ref{A'inter}), we obtain
\begin{align}
\langle \mathcal{A'}_{\mathcal{S}\mathcal{E}}(q_k, \circledcirc) \rangle_{\circledcirc | q_k}
&\nonumber= \int_{V_\mathcal{E}} \mathrm{d}q  \sum_{\substack{j=1 \\ j \neq k}}^{N} 
\mathcal{L}_{q_k}^{\circledcirc} \left\{ P(\circledcirc \mid q_k)\, \delta(q - q_j) \right\}
\,\\&\times u\big( \| q - q_k \|_2 \big).
\label{A_exp}
\end{align}
To evaluate the term $\mathcal{L}_{q_k}^{N} \{ P(\circledcirc \mid q_k) \delta(q - q_j) \}$ in the integrand, we apply the general identity introduced in Eq.~(\ref{eq:reduced_cond_expectation}). 
In addition, by relating the generic and specific reduced distributions (see Appendix~\ref{$n$-Particle Densities} and \ref{Derivation of the first moment of Interaction}), we can write
\begin{align}
\langle \mathcal{A'}_{\mathcal{S}\mathcal{E}}(q_k, \circledcirc) \rangle_{\circledcirc \mid q_k} 
= \int_{V_\mathcal{E}} \mathrm{d}q \, \rho(q)\, g^{(2)}(q, q_k)\, u\big( \| q - q_k \|_2 \big),
\end{align}
where $\rho(\bullet)$ is the generic single-particle density and $g^{(2)}(\bullet, \bullet)$ denotes the two-body correlation function.
Finally, substituting this result back into Eq.~(\ref{V_expect}), we obtain  
\begin{align}
\langle V_{\mathcal{S}\mathcal{E}} \rangle
= \int_{V_\mathcal{S}} \mathrm{d}q_\odot \, \rho(q_\odot) \int_{V_\mathcal{E}} \mathrm{d}q \, \rho(q) g^{(2)}(q, q_\odot) u(\| q - q_\odot \|_2). \label{eq:Expectation}  
\end{align}  
This result provides a clear way to compute the average interaction energy using structural properties. 
Eq.~(\ref{eq:Expectation}) provides an averaged interaction energy, incorporating not only the spatial structure encoded by $\rho(\bullet)$ but also many-body correlations captured by $g^{(2)}(\bullet, \bullet)$. 
In Appendix \ref{Derivation of the first moment of Interaction} we provide a more detailed derivation of this first moment of the system-environment interaction energy based on our reference particle approach.

\section{Derivation of the Second Moment of Interaction }\label{DSM}
Next, we derive an expression for the second moment of the interaction, defined as the expectation value of the squared interaction energy, $\langle V_{\mathcal{S}\mathcal{E}}^2 \rangle$. 
Following the procedure of the previous section, and again starting from Eq.~(\ref{eq:interaction_expectation_initial}), we obtain
\begin{equation}
\langle V^2_{\mathcal{S}\mathcal{E}} \rangle = \frac{\mathcal{L}^{N} \{ \delta \mathcal{C}(x_\mathcal{E}, x_\mathcal{S}; E) \,V^2_{\mathcal{S}\mathcal{E}}(x_{\mathcal{E}}, x_{\mathcal{S}})\}}{\Omega}.
\label{eq:interaction_deviation}
\end{equation}
Focusing on interactions that can be expressed as a sum over pairwise terms, we can write
\begin{equation}
V^2_{\mathcal{S}\mathcal{E}}(x_{\mathcal{E}}, x_{\mathcal{S}}) =  \left(\sum_{k \in \mathcal{I}{\{\mathcal{S}\}(x)}}   \sum_{j \in \mathcal{I}{\{\mathcal{E}\}(x)}} u(\| q_k - q_j \|_2)\right)^2.
\label{power2}
\end{equation}
To arrive at a computable expression, we reformulate the equation above using the reference particle approach. 
Specifically, the squared interaction energy can be represented in the form
\begin{equation}
V^2_{\mathcal{S}\mathcal{E}}(x_{\mathcal{E}}, x_{\mathcal{S}}) = \sum_{k \in \mathcal{I}\{\mathcal{S}\}(x)} \mathcal{A'}_{\mathcal{S}\mathcal{E}}(q_k, \circledcirc).
\label{decom}
\end{equation}
The central task now is to determine the explicit mathematical form of $\mathcal{A'}_{\mathcal{S}\mathcal{E}}(q_k, \circledcirc)$ that appears in the second-moment decomposition.
In the case of the first moment, this form was directly inferred by comparing Eq.~(\ref{eq:three}) with Eq.~(\ref{eq:four}).
While the derivation is slightly more involved for the second moment, we adopt the same decomposition logic to extract its structure. Instead of algebraically expanding all terms to identify those that involve the reference particle $k$, we rely on a schematic representation of the pairwise interaction terms. This graphical approach helps isolate all contributions involving $k$ and organize them systematically. 

\begin{figure}[htbp]
    \centering    \includegraphics[width=\columnwidth]{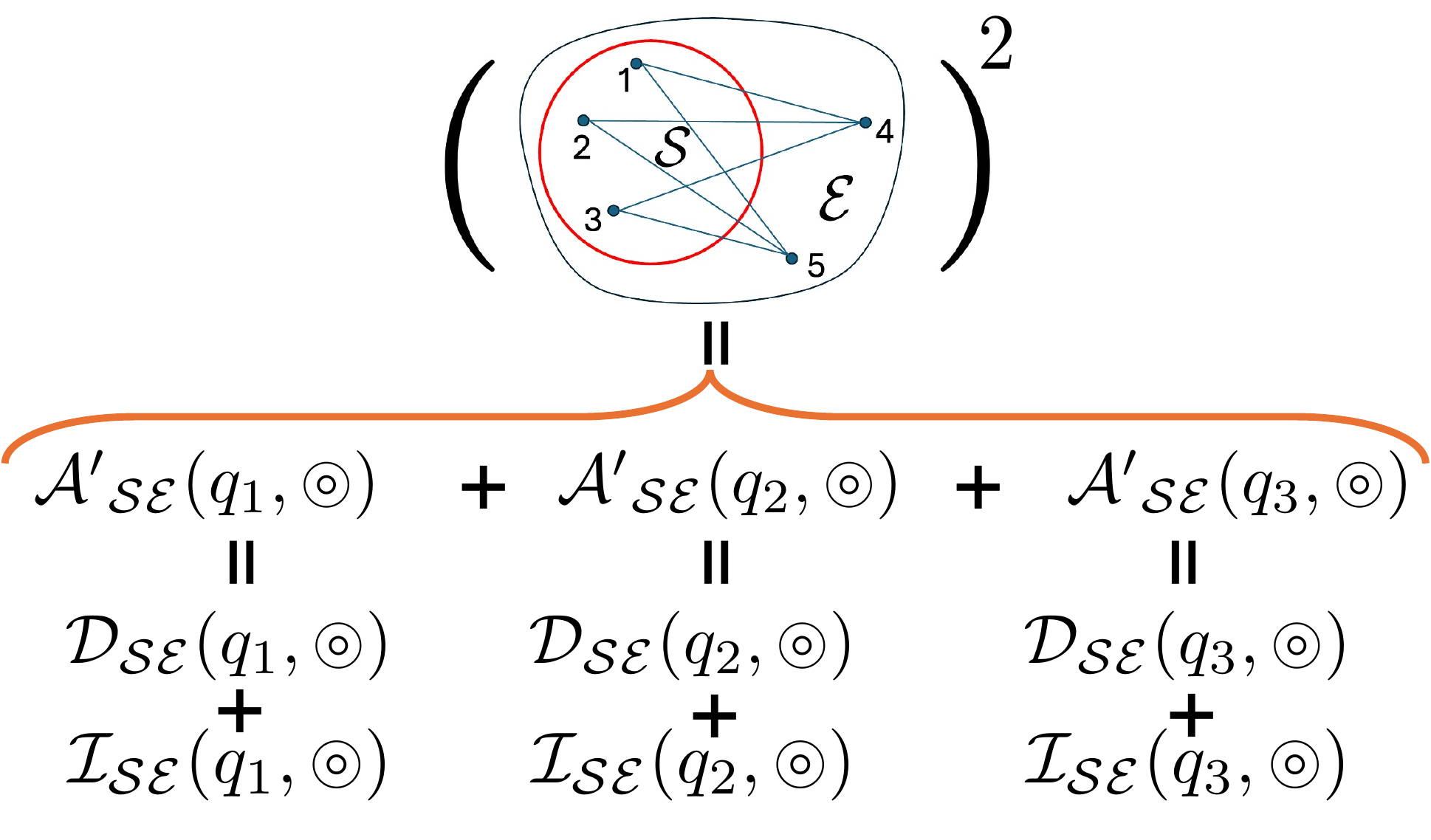}
    \caption{Graphical representation of the decomposition of the squared interaction $ V^2_{\mathcal{S}\mathcal{E}}$ into contributions from each reference particle (see Eq.~(\ref{decom})). Each contribution, $A'_{\mathcal{S}\mathcal{E}}(q_k, \circledcirc)$, is further divided into a diagonal $\mathcal{D}_{\mathcal{S}\mathcal{E}}$ and an off-diagonal $\mathcal{I}_{\mathcal{S}\mathcal{E}}$ component (see Eq.~(\ref{do})).}
    \label{fig:decomposition}
\end{figure}

\begin{figure}[htbp]
    \centering
    \includegraphics[width=\columnwidth]{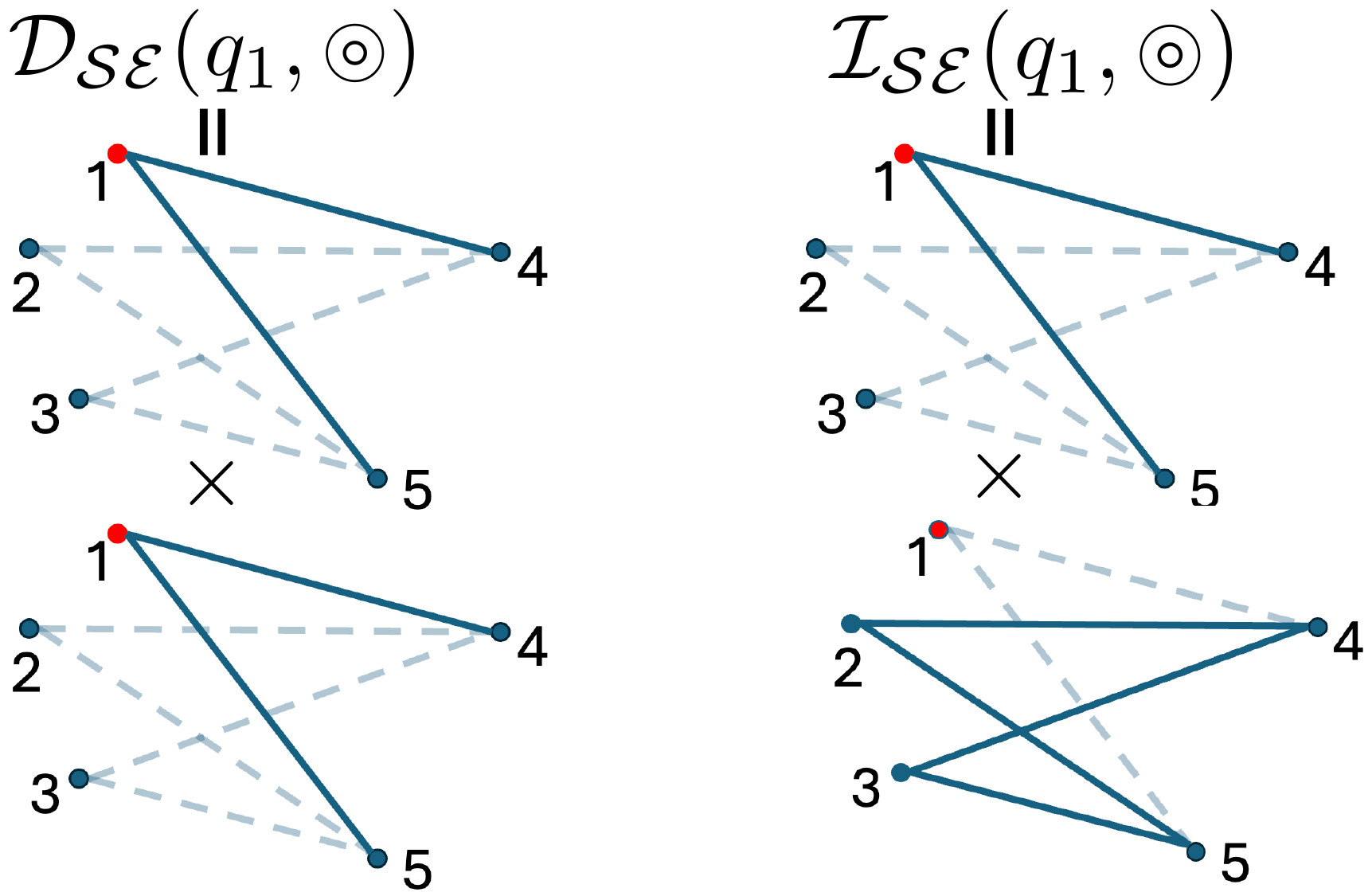}
    \caption{Illustration of the diagonal $\mathcal{D}_{\mathcal{S}\mathcal{E}}(q_1, \circledcirc)$ and off-diagonal $\mathcal{I}_{\mathcal{S}\mathcal{E}}(q_1, \circledcirc)$ components of the interaction term $A'_{\mathcal{S}\mathcal{E}}(q_1, \circledcirc)$, using reference particle $q_1$ as an example.}
\label{fig:refparticle1}
\end{figure}

As shown in Fig.~\ref{fig:decomposition}, the total contribution associated with particle $k$ can be partitioned into two parts: a diagonal term, denoted $\mathcal{D}_{\mathcal{S}\mathcal{E}}(q_k, \circledcirc)$, and an off-diagonal term, $\mathcal{I}_{\mathcal{S}\mathcal{E}}(q_k, \circledcirc)$.  
To illustrate this, we consider the same five-particle system introduced earlier.
By explicitly expanding the square of the total interaction energy in Eq.~(\ref{power2}), one finds that the contribution involving particle $k=1$ takes the form
\begin{align}
&\nonumber\mathcal{D}_{\mathcal{S}\mathcal{E}}(q_1, \circledcirc) = (u_{14} + u_{15})^2, \qquad \\&
\mathcal{I}_{\mathcal{S}\mathcal{E}}(q_1, \circledcirc) = (u_{14} + u_{15})(u_{24} + u_{25} + u_{34} + u_{35}),
\end{align}
as shown in Fig.~\ref{fig:refparticle1}.
In general, the diagonal term corresponds to the square of the sum of all pairwise interactions between the reference particle and the environment. 
The off-diagonal term captures the cross-contributions between the reference particle and all other system particles, each interacting with the environment. 
This classification allows us to conclude that
\begin{align}
&\nonumber\mathcal{A'}_{\mathcal{S}\mathcal{E}}(q_k, \circledcirc) \\&\nonumber= \underbrace{\sum_{j \in \mathcal{I}{\{\mathcal{E}\}(x)}} \sum_{j' \in \mathcal{I}{\{\mathcal{E}\}(x)}}  u(\lVert q_j - q_k \rVert_2 ) u(\lVert q_{j'} - q_k \rVert_2 )}_{\mathcal{D}_{\mathcal{S}\mathcal{E}}(q_k, \circledcirc )}\\& +\underbrace{ \sum_{\substack{* \in \mathcal{I}{\{\mathcal{S}\}(x)} \\ *\neq k }} \sum_{j \in \mathcal{I}{\{\mathcal{E}\}(x)}} \sum_{j' \in \mathcal{I}{\{\mathcal{E}\}(x)}} u(\lVert q_j - q_k \rVert_2 ) u(\lVert q_{j'} - q_* \rVert_2 )}_{\mathcal{I}_{\mathcal{S}\mathcal{E}}(q_k, \circledcirc)}.
\label{do}
\end{align}
Using the transformation identity developed in Eq.~(\ref{powertrans}), the diagonal and off-diagonal contributions can be written explicitly as
\begin{align}
\mathcal{D}_{\mathcal{S}\mathcal{E}}(q_k, \circledcirc )  &\nonumber=\int_{V_\mathcal{E}} \mathrm{d}q \int_{V_\mathcal{E}} \mathrm{d}q'   \sum_{\substack{j=1\\ j\neq k}}^{N} \sum_{\substack{j'=1\\ j'\neq k} }^{N}   \\&\nonumber \times \delta(q - q_j) \delta(q' - q_{j'})u(\lVert q - q_k \rVert_2 ) \\&\times u(\lVert q' - q_k \rVert_2).
\label{D}
\end{align}
and
\begin{align}
\mathcal{I}_{\mathcal{S}\mathcal{E}}(q_k, \circledcirc)&\nonumber= \int_{V_\mathcal{S}} \mathrm{d}q_* \int_{V_\mathcal{E}} \mathrm{d}q \int_{V_\mathcal{E}} \mathrm{d}q'  \sum_{\substack{j=1\\ j\neq k}}^{N} \sum_{\substack{j'=1\\ j'\neq k}}^{N} \sum_{\substack{j''=1\\ j''\neq k}}^{N}\\&\times\nonumber  \delta(q - q_j) \delta(q' - q_{j'}) \delta(q_* - q_{j''}) \\&\times u(\lVert q - q_k \rVert_2 ) u(\lVert q' - q_* \rVert_2 ).
\label{I}
\end{align}
Plugging these two expressions for the diagonal and off-diagonal parts into Eq.~(\ref{central}) and employing the linearity of the expectation operator over summation, we obtain 
\begin{align}
\langle V^2_{\mathcal{S}\mathcal{E}} \rangle &\nonumber= \underbrace{\int_{V_\mathcal{S}} \mathrm{d}q_\odot \, \rho(q_\odot) \delta_{q_\odot{q}_{k}} \otimes  \langle \mathcal{D}_{\mathcal{S}\mathcal{E}}(q_k, \circledcirc )\rangle_{\circledcirc|q_k}}_{\mathcal{D}}\\&+\underbrace{\int_{V_\mathcal{S}} \mathrm{d}q_k \, \rho(q_\odot) \delta_{q_\odot{q}_{k}} \otimes  \langle \mathcal{I}_{\mathcal{S}\mathcal{E}}(q_k, \circledcirc)\rangle_{\circledcirc|q_k}}_{\mathcal{I}},
\end{align}
where we will refer to $\mathcal{D}$ and $\mathcal{I}$  as the diagonal and off-diagonal integral, respectively. 
We begin by evaluating the diagonal integral.
The relevant term is the conditional expectation
$\langle \mathcal{D}_{\mathcal{S}\mathcal{E}}(q_k, \circledcirc ) \rangle_{\circledcirc | q_k}$, which, using Eq.~(\ref{D}) and Eq.~(\ref{eq:single_particle_cond_exp}), can be written as
\begin{align}
\langle \mathcal{D}_{\mathcal{S}\mathcal{E}}(q_k, \circledcirc )\rangle_{\circledcirc|q_k} &\nonumber= \int_{V_\mathcal{E}} \mathrm{d}q \int_{V_\mathcal{E}} \mathrm{d}q'  \sum_{\substack{j=1\\ j\neq k}}^{N} \sum_{\substack{j'=1\\ j'\neq k}}^{N}\\&\nonumber\times\mathcal{L}_{q_k}^{N} \{P(\circledcirc|q_k)\delta(q - q_j) \delta(q' - q_{j'})\} \\&\times u(\lVert q - q_k \rVert_2) u(\lVert q' - q_k \rVert_2).
\end{align}
After evaluating the expressions above (see Appendix~\ref{Derivation of the Second Moment of Interaction} for a detailed derivation) and substituting the result into the diagonal integral, we find
\begin{equation}
\mathcal{D} = M + R,
\end{equation}
where $M$ captures contributions involving two-body correlations:
\begin{align}
M = \int_{V_\mathcal{S}} \mathrm{d}q_\odot \, \rho(q_\odot) \int_{V_\mathcal{E}} \mathrm{d}q \, \rho(q) \, g^{(2)}(q, q_\odot)\, u^2(\lVert q - q_\odot \rVert_2),
\label{eq:M_term}
\end{align}
and $R$ accounts for the three-body correlation function:
\begin{align}
R &= \int_{V_\mathcal{S}} \mathrm{d}q_\odot \, \rho(q_\odot) \int_{V_\mathcal{E}} \mathrm{d}q' \, \rho(q') \int_{V_\mathcal{E}} \mathrm{d}q \, \rho(q) \nonumber\\
&\quad \times g^{(3)}(q_\odot, q, q')\, u(\lVert q - q_\odot \rVert_2 )\, u(\lVert q' - q_\odot \rVert_2 ).
\label{eq:R_term}
\end{align}
Here, $\rho(\bullet)$ denotes the single-particle density, while $g^{(2)}(\bullet, \bullet)$ and $g^{(3)}(\bullet, \bullet, \bullet)$ are the standard two-body and three-body correlation functions, respectively. 
Turning now to the off-diagonal contribution, we proceed analogously by computing the conditional expectation $\langle \mathcal{I}_{\mathcal{S}\mathcal{E}}(q_k, \circledcirc)\rangle_{\circledcirc|q_k}$. 
Using Eq.~(\ref{I}) in conjunction with Eq.~(\ref{eq:single_particle_cond_exp}), we obtain
\begin{align}
\langle \mathcal{I}_{\mathcal{S}\mathcal{E}}(q_k, \circledcirc) \rangle_{\circledcirc|q_k} &= \int_{V_\mathcal{S}} \mathrm{d}q_* \int_{V_\mathcal{E}} \mathrm{d}q \int_{V_\mathcal{E}} \mathrm{d}q' \nonumber\\
&\quad \times \sum_{\substack{j=1\\ j\neq k}}^{N} \sum_{\substack{j'=1\\ j'\neq k}}^{N} \sum_{\substack{j''=1\\ j''\neq k}}^{N} \mathcal{L}_{q_k}^{N} \left\{ P(\circledcirc \mid q_k) \right. \nonumber\\
&\quad \left. \times \delta(q - q_j)\, \delta(q' - q_{j'})\, \delta(q_* - q_{j''}) \right\} \nonumber\\
&\quad \times u(\lVert q - q_k \rVert_2) \, u(\lVert q' - q_* \rVert_2).
\end{align}
After evaluating this expression (see Appendix~\ref{Derivation of the Second Moment of Interaction} for detailed steps) and substituting it into the off-diagonal integral, we obtain the decomposition
\begin{equation}
\mathcal{I} = C + S,
\end{equation}
where $C$ and $S$ include three-body and four-body correlation terms, respectively.
Specifically, the first term $C$ is given as
\begin{align}
C &= \int_{V_\mathcal{S}} \mathrm{d}q_\odot \, \rho(q_\odot) \int_{V_\mathcal{S}} \mathrm{d}q_* \, \rho(q_*) \int_{V_\mathcal{E}} \mathrm{d}q \, \rho(q) \nonumber \\
&\quad \times g^{(3)}(q_\odot, q, q_*) \, u(\lVert q - q_\odot \rVert_2 ) \, u(\lVert q - q_* \rVert_2 ),
\label{eq:C_term}
\end{align}
and the second term $S$ can be written as
\begin{align}
S &= \int_{V_\mathcal{S}} \mathrm{d}q_\odot \, \rho(q_\odot) \int_{V_\mathcal{S}} \mathrm{d}q_* \, \rho(q_*) \int_{V_\mathcal{E}} \mathrm{d}q' \, \rho(q') \int_{V_\mathcal{E}} \mathrm{d}q \, \rho(q) \nonumber \\
&\quad \times g^{(4)}(q_\odot, q, q', q_*) \, u(\lVert q - q_\odot \rVert_2 ) \, u(\lVert q' - q_* \rVert_2 ).
\label{eq:J_term}
\end{align}
Here, $g^{(4)}(\bullet, \bullet, \bullet, \bullet)$ denotes four-body correlation functions. Together with the diagonal contributions in Eqs.~(\ref{eq:M_term}–\ref{eq:R_term}), this completes the decomposition of the second moment $\langle V_{\mathcal{S}\mathcal{E}}^2 \rangle$ into physically interpretable components as
\begin{equation}
\langle V_{\mathcal{S}\mathcal{E}}^2 \rangle =  M + R + C + S \, .
\label{second_moment}
\end{equation}

\section{Connection to $\Delta \mathcal{F}_{\mathcal{S}} = \mathcal{F}^*_{\mathcal{S}} - \mathcal{F}_{\mathcal{S}}$}
\label{secfree}
\begin{figure}[htbp]
    \centering
    \includegraphics[width=\columnwidth]{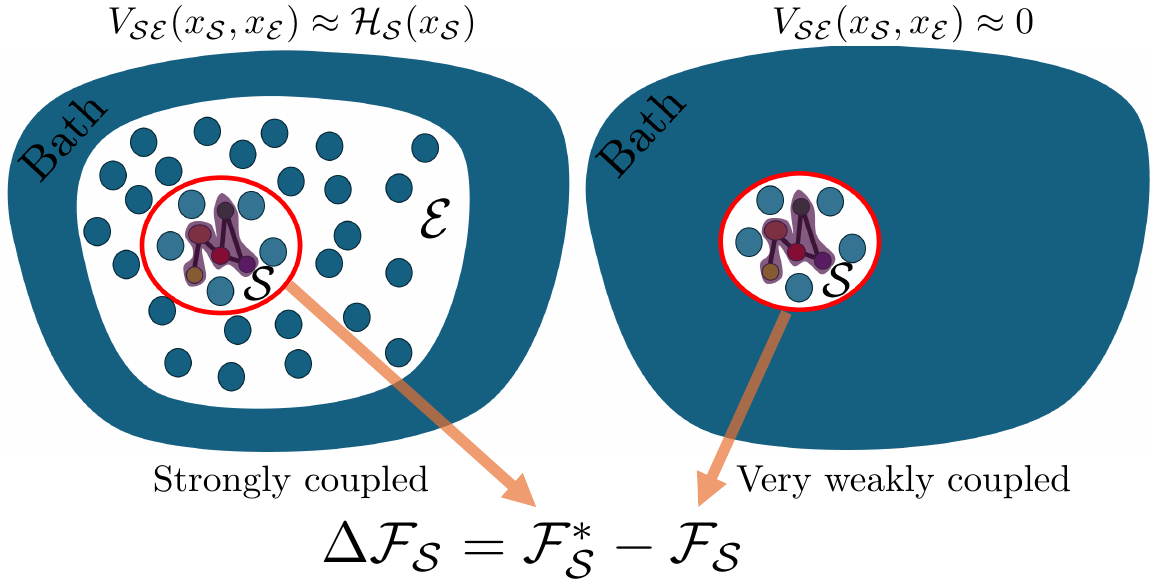}
    \caption{Free‑energy difference $\Delta \mathcal{F}_{\mathcal{S}}= \mathcal{F}^*_{\mathcal{S}} - \mathcal{F}_{\mathcal{S}}$ between the  strong‑coupling limit ($V_\mathcal{SE}\approx \mathcal{H}_\mathcal{S}$) and the weak‑coupling limit ($V_\mathcal{SE}\approx0$).}
    \label{F_dif}
\end{figure}

In this section, we demonstrate how the results developed within the present framework can be applied to analyze systems under strong coupling. 
In our companion article \cite{PRL}, we discuss in detail how the distribution of interaction energies, $P(V_{\mathcal{SE}})$, plays a central role for strongly coupled systems.
Here, we focus specifically on evaluating the free energy difference between a strongly coupled system and its very weakly coupled counterpart, with both states assumed to be in thermal equilibrium with the same heat bath (see Fig.~\ref{F_dif}). 

In the HMF formalism \cite{burke2024structure,talkner2016open,talkner2020colloquium}, the effective free energy of the system $\mathcal{S}$ in contact with an environment $\mathcal{E}$ is defined by
\begin{equation}
\mathcal{F}^*_{\mathcal{S}}(\beta) = -\beta^{-1} \ln \mathcal{Z}_*
\end{equation}
where
\begin{equation}
\mathcal{Z}_*(\beta) = \frac{\mathcal{Z}_{\mathcal{S}+\mathcal{E}}(\beta)}{\mathcal{Z}_{\mathcal{E}}(\beta)}.
\end{equation}
This particular form of the partition function provides a consistent thermodynamic description of the open system, regardless of the strength of its interaction with the environment \cite{talkner2020colloquium}.
Using the definition above, we can write
\begin{align}
\mathcal{F}^*_{\mathcal{S}}(\beta) 
= -\beta^{-1} \ln \mathcal{Z}_{\mathcal{S}+\mathcal{E}} + \beta^{-1} \ln \mathcal{Z}_{\mathcal{E}} 
= \mathcal{F}_{\mathcal{S}+\mathcal{E}} - \mathcal{F}_{\mathcal{E}}.
\label{free energy diff1}
\end{align}
The effect of coupling leads to the free energy difference
\begin{equation}
\Delta \mathcal{F}_{\mathcal{S}}\equiv \mathcal{F}^*_{\mathcal{S}} - \mathcal{F}_{\mathcal{S}}\, .
\label{free energy diff2}
\end{equation}
By substituting Eq.~\ref{free energy diff2} into Eq.~\ref{free energy diff1}, we obtain the following result
\begin{equation}
\Delta \mathcal{F}_{\mathcal{S}}= \mathcal{F}_{\mathcal{S}+\mathcal{E}} - \mathcal{F}_{\mathcal{E}} - \mathcal{F}_{\mathcal{S}}.
\label{free energy diff3}
\end{equation}
More precisely, the free energy shift $\Delta \mathcal{F}_{\mathcal{S}}$ can be expressed as the difference between $F_{\mathcal{S+E}}$, the free energy of the fully coupled system and environment in equilibrium with a thermal bath, and the sum $F_\mathcal{S} + F_\mathcal{E}$, which corresponds to the case where system and environment are completely decoupled and independently equilibrated. 
This viewpoint naturally aligns with the FEP formalism, where the free energy difference between a reference state and a target state is given by \cite{chipot2007free,zwanzig1954high}
\begin{equation}
e^{-\beta (\mathcal{F}_{\mathcal{S+E}} - \mathcal{F}_\mathcal{S} - \mathcal{F}_\mathcal{E})}
= \Bigl\langle e^{-\beta V_{\mathcal{SE}}} \Bigr\rangle_{\mathcal{S}_0 + \mathcal{E}_0}
= \Bigl\langle e^{\beta V_{\mathcal{SE}}} \Bigr\rangle_{\mathcal{S} + \mathcal{E}}^{-1},
\label{FEP_0}
\end{equation}
where the averages on the right-hand side are taken over the uncoupled and fully coupled ensembles, respectively.
If the interaction energies $V_{\mathcal{SE}}$ in both the coupled and uncoupled ensembles are approximately Gaussian, then the corresponding exponential averages can be evaluated analytically. 
In particular, one finds
\begin{align}
\Delta \mathcal{F}_{\mathcal{S}} =  \mu_0 - \frac{1}{2} \beta \sigma_0^2
\label{ref}
\end{align}
or
\begin{align}
\Delta \mathcal{F}_{\mathcal{S}} = \mu + \frac{1}{2} \beta \sigma^2
\label{tar}
\end{align}
where $\mu_0$ and $\sigma_0^2$ are the mean and variance of $V_{\mathcal{SE}}$ in the uncoupled ensemble, while $\mu$ and $\sigma^2$ refer to the same quantities in the coupled case.

Before proceeding, it is important to highlight a key conceptual point. 
As illustrated in Fig.~\ref{F_dif}, the system under consideration consists of a fixed number of particles, analyzed in two distinct scenarios: one in which the system is non-interacting (or very weakly coupled) with its environment, and another in which it is strongly coupled.
In the strongly coupled case, we assume that there is no exchange of particles with the surroundings, and only energy transfer is permitted. 
This assumption is intrinsic to the HMF framework, which adopts a canonical viewpoint for the system \cite{jarzynski2017stochastic,talkner2020colloquium}. 
By contrast, in this work, we considered a more general setting in which the system --- more precisely, the region of interest --- can exchange both energy and particles with the environment. 
In other words, the results derived here apply to a broader class of open systems, with the fixed-particle-number setup treated in Fig.~\ref{F_dif} representing a specific case of that general framework. 
We now clarify how the proposed framework can be used to compute the free energy shift $\Delta \mathcal{F}_{\mathcal{S}}$.
Starting from Eq.~(\ref{ref}), the mean interaction energy $\mu_0$ in the non-interacting scenario is given by
\begin{align}
\mu_0
= \int_{V_\mathcal{S}} \mathrm{d}q_\odot \, \rho(q_\odot) \int_{V_\mathcal{E}} \mathrm{d}q \, \rho(q)\, g^{(2)}(q, q_\odot)\, u(\| q - q_\odot \|_2).
\end{align}
The corresponding variance takes the form
\begin{equation}
\sigma_0^2 = M + R + C + S - \mu_0^2\, .
\end{equation}
A key point is that all inputs appearing in the equations above ---$\rho(\bullet)$, $g^{(2)}(\bullet, \bullet)$, $g^{(3)}(\bullet, \bullet, \bullet)$, and $g^{(4)}(\bullet, \bullet, \bullet, \bullet)$--- must be computed under the assumption that the system and environment are decoupled. 
These quantities must therefore be obtained by sampling configurations in which the full composite supersystem remains in thermal equilibrium with the external bath, while no interaction is present between the system and the environment. 
By contrast, if one uses Eq.~(\ref{tar}), the same expressions for $\mu$ and $\sigma$ apply, but the statistical quantities must be evaluated in the fully interacting ensemble. 
This distinction reflects the structural flexibility of Eqs.~(\ref{eq:Expectation}) and (\ref{second_moment}), which, like the FEP method, support dual interpretations based on either reference or target statistics. This property underlines the generality of the approach developed here.

\section{Model Validation}
\label{secmode}

To validate our approach, we consider a collection of identical particles confined within a box. 
The system of interest, $\mathcal{S}$, is defined as a spherical region centered inside the box, while the surrounding space constitutes the environment, $\mathcal{E}$.
Particles are allowed to exchange between the system and its environment, and energy transfer is also permitted.
The entire composite system --- comprising $\mathcal{S}$ and $\mathcal{E}$ --- is modeled within the canonical ($NVT$) ensemble, maintaining fixed total particle number $N$, volume $V$, and temperature $T$. 
In the scenario where the system $\mathcal{S}$ is large, interactions with the environment $\mathcal{E}$ occur primarily at the two-dimensional surface, involving only a small fraction of the system's particles \cite{jarzynski2017stochastic}. 
Consequently, the interaction energy $V_{\mathcal{S}\mathcal{E}}$ is negligible compared to the internal energy $\mathcal{H}_{\mathcal{S}}$, allowing us to clearly identify weak coupling. 
In this regime, the Hamiltonian simplifies to $\mathcal{H}_{\mathcal{E} + \mathcal{S}} = \mathcal{H}_{\mathcal{E}} + \mathcal{H}_{\mathcal{S}}$. For small systems, however, this distinction blurs. The "surface" may encompass most or all of the system's degrees of freedom, making $V_{\mathcal{S}\mathcal{E}}$ comparable in magnitude to $\mathcal{H}_{\mathcal{S}}$. 
In such cases, the interaction term cannot be neglected, and the Hamiltonian of the combined system becomes $\mathcal{H}_{\mathcal{E} + \mathcal{S}} = \mathcal{H}_{\mathcal{E}} + \mathcal{H}_{\mathcal{S}} + V_{\mathcal{S}\mathcal{E}}$\cite{jarzynski2017stochastic,talkner2020colloquium,campisi2009fluctuation}. 
We quantify the internal energy $\langle \mathcal{H}_{\mathcal{S}} \rangle$ and the interaction energy $\langle V_{\mathcal{S}\mathcal{E}} \rangle$, recognizing both as extensive variables proportional to the system size. 
Specifically, they are proportional to the average number of particles in the system, $\langle N_{\mathcal{S}} \rangle$, and the average number of particles contributing to the interaction, $\langle N_{\mathcal{S}\mathcal{E}} \rangle$, respectively.
By modeling the combined system as a large box and $\mathcal{S}$ as a sphere in its center with $\mathcal{E}$ occupying the surrounding space, we conclude that $\langle N_{\mathcal{S}} \rangle \propto R_{\mathcal{S}}^3$ and $\langle N_{\mathcal{S}\mathcal{E}} \rangle \propto R_{\mathcal{S}}^2$, where $R_{\mathcal{S}}$ is the sphere's radius. 
This leads to $\langle V_{\mathcal{S}\mathcal{E}} \rangle \propto R_{\mathcal{S}}^2$ and $\langle (V_{\mathcal{S}\mathcal{E}} - \langle V_{\mathcal{S}\mathcal{E}} \rangle)^2 \rangle \propto R_{\mathcal{S}}^2$. 

Our model expresses the probability distribution of a system observable $x_{\mathcal{S}}$ as $P(x_{\mathcal{S}}) = \int \mathrm{d}(V_{\mathcal{S}\mathcal{E}}/\langle N_{\mathcal{S}} \rangle) P(x_{\mathcal{S}} | V_{\mathcal{S}\mathcal{E}}/\langle N_{\mathcal{S}} \rangle) \mathcal{N}(\alpha, \gamma)$, where $\alpha = \langle V_{\mathcal{S}\mathcal{E}}/\langle N_{\mathcal{S}} \rangle \rangle$ and $\gamma = \langle (V_{\mathcal{S}\mathcal{E}}/\langle N_{\mathcal{S}} \rangle - \alpha)^2 \rangle$. 
As the system size increases ($R_{\mathcal{S}} \rightarrow R_{\text{TL}}$), approaching the thermodynamic limit, both $\alpha \propto 1/R_{\mathcal{S}}$ and $\gamma \propto 1/R_{\mathcal{S}}^4$ approach zero. Consequently, $P(V_{\mathcal{S}\mathcal{E}}/\langle N_{\mathcal{S}} \rangle)$ converges to a Dirac delta function, and the distribution simplifies to $P(x_{\mathcal{S}}) = P(x_{\mathcal{S}} |(V_{\mathcal{S}\mathcal{E}}/\langle N_{\mathcal{S}} \rangle) = 0)$, reinforcing the assumption of weak coupling. 
To assess the model in the strong coupling regime, we performed Monte Carlo simulations of the combined system $\mathcal{S} + \mathcal{E}$ within an $NVT$ ensemble, using particles interacting via Lennard-Jones potentials. 
Utilizing reduced units normalized by the Lennard-Jones parameters $\sigma$ and $\epsilon$, we sampled configurations and calculated the interaction energy $V_{\mathcal{S}\mathcal{E}}$.
This computationally intensive approach provided reference data with the combined system fully modeled in explicit detail. 
We then computed the expected value and variance of $V_{\mathcal{S}\mathcal{E}}$ using our method, confirming that it reproduces the reference distribution. 
\begin{figure}[htbp]
\includegraphics[width=\columnwidth]{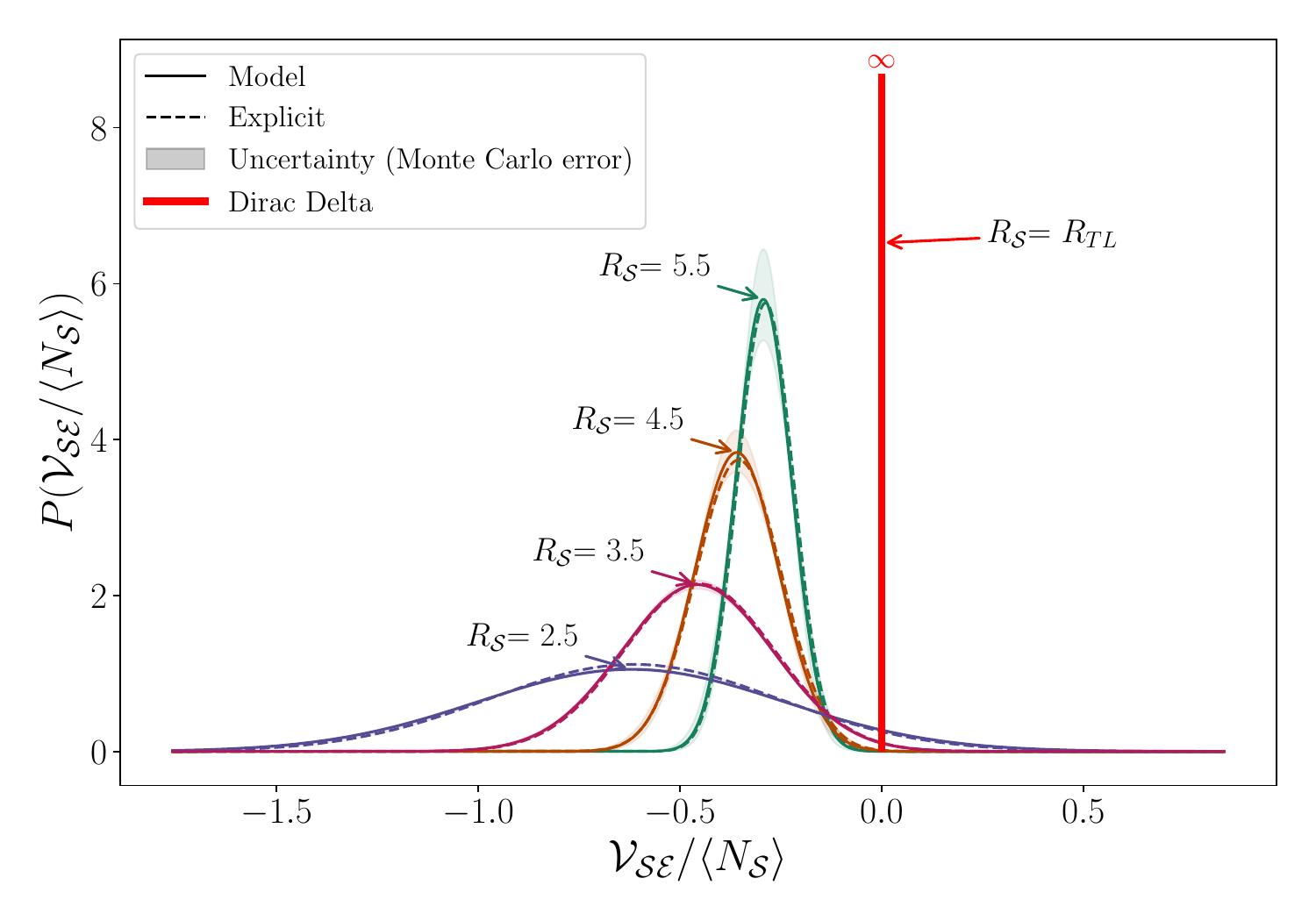}
\caption{The probability distribution of the normalized interaction energy, $V_{\mathcal{S}\mathcal{E}} / \langle N_S \rangle$, for various system sizes ($R_S$). As $R_S$ increases, the distribution progressively narrows, transitioning from a broad profile indicative of a strong coupling regime at small sizes to a Dirac delta function characteristic of a weak coupling regime in the thermodynamic limit. We compare simulation results (Explicit) with theoretical predictions (Model), showing excellent agreement and validating the accuracy of the proposed framework across different coupling regimes.} \label{fig:normalized interaction}
\end{figure}

Figure \ref{fig:normalized interaction} shows the probability distribution as a function of the normalized interaction energy, $V_{\mathcal{S}\mathcal{E}} / \langle N_S \rangle$, for different system sizes.
The numerical results confirm that for small system sizes (small $R_S$), the distribution is broad, indicating significant fluctuations in the interaction energy, which corresponds to a strong coupling regime. As the system size increases, the distribution of the interaction energy narrows as the variance of the interaction energy decreases. 
In the thermodynamic limit (as $R_S$ approaches $R_{TL}$), the distribution converges to a Dirac delta function and the system is in a weak coupling regime. 
The excellent agreement between results obtained with our proposed framework (Model) and the expensive $NVT$ Monte Carlo simulations (Explicit) validates the model's accuracy in capturing the behavior of open systems across different coupling regimes. 
\begin{figure}[htbp]
    \centering
\includegraphics[width=\columnwidth]{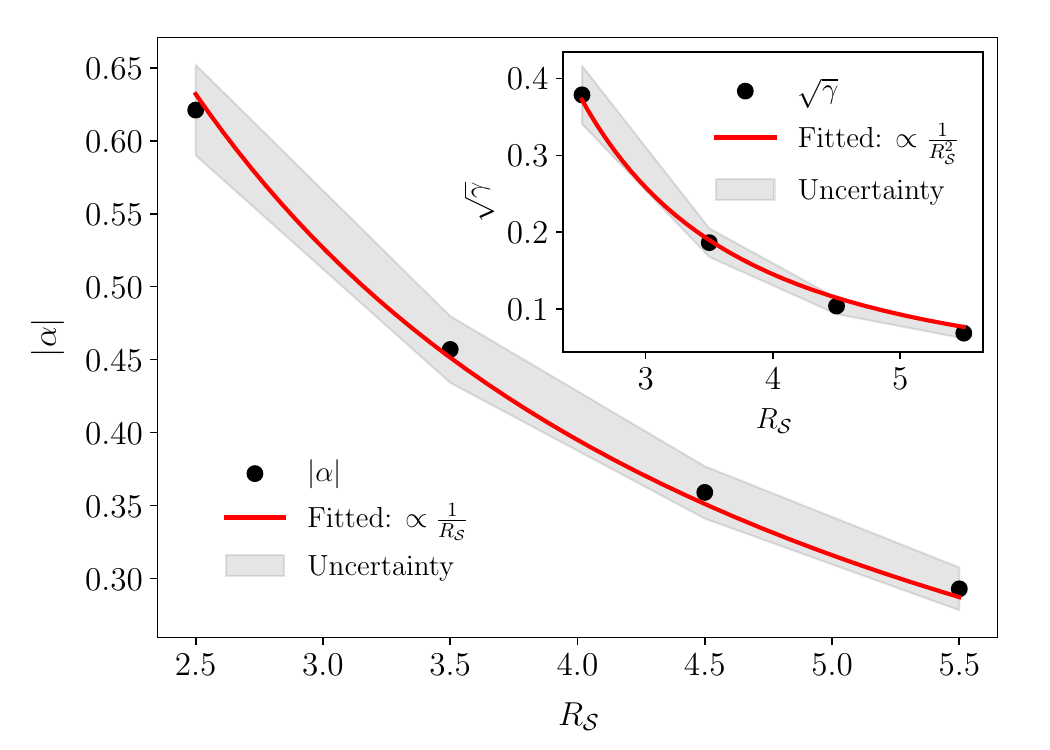}
    \caption{Mean ($|\alpha|$) and variance ($\sqrt{\gamma}$) of the normalized interaction energy distribution as a function of system size ($R_S$). The main graph shows that $|\alpha|$ decreases proportionally to $1/R_S$, while the inset demonstrates that $\sqrt{\gamma}$ decreases proportionally to $1/R_S^2$. These results demonstrate that both the mean and fluctuations of the interaction energy become negligible as the system size increases, in agreement with the theoretical prediction.}
    \label{fig:system size}
\end{figure}
In Figure \ref{fig:system size} we quantify the relationship between system size and the first and second moments of the interaction energy: The main graph shows the absolute value of the normalized expectation of the interaction, $|\alpha|$, as a function of the system size, $R_S$. 
Clearly, as $R_S$ increases, $|\alpha|$ decreases and approaches zero. 
The red fit curve demonstrates the inverse proportionality with $R_S$, consistent with theoretical predictions. 
The inset shows the square root of the variance of the normalized interaction, $\sqrt{\gamma}$, as a function of the system size, $R_S$. As $R_S$ increases, also $\sqrt{\gamma}$ decreases and approaches zero. 
The fit demonstrates a proportionality with $1/R_S^2$, which is also consistent with theoretical predictions.  
This means that the interaction energy becomes negligible compared to the system’s internal energy. 
The calculation of the expectation and the variance in Eqs.~(\ref{eq:Expectation}) and ~(\ref{second_moment}), requires to solve high-dimensional integrals (up to twelve dimensions). 
We employed Monte Carlo integration and utilized established approximations for low-density systems to simplify the evaluation of pair, triple, and quadruple correlation functions, as well as local density distributions.
Specifically, we used the Boltzmann factor for the radial distribution function \cite{hansen2013theory}, Kirkwood’s superposition approximation for the triple correlation function \cite{kirkwood1935statistical}, and the Fisher-Kopeliovich closure for the quadruple correlation function \cite{fisher1960refinement, meeron1957series, salpeter1958mayer,grouba2004superposition}. 
Our numerical results confirm the validity of the proposed model across different coupling regimes. The model not only aligns with thermodynamic behavior in the thermodynamic limit --- where increasing $R_{\mathcal{S}}$ causes the distribution to approach a Dirac delta function --- but also provides accurate predictions for small systems with strong coupling, as evidenced by the close match between the distributions obtained with our newly proposed framework and those obtained from the Monte Carlo sampled reference simulations. 

We further note that our key motivation was to split the combined system into an active system and an environment. We can create a limiting case of our theory, where we remove that distinction such that we have only one system of uniform density. If we further treat the Lennard-Jones potential as a perturbation to a hard-sphere model, our derived first and second moments of the interaction collapse to Eqs.~(20) and (25) in Robert W. Zwanzig's original paper on free energy perturbation \cite{zwanzig1954high}.

\section{Conclusion}
\label{seccon}
We have shown that by treating the system–environment interaction energy $V_{\mathcal{SE}}$ as a random variable and applying the reference-particle decomposition, one obtains exact integral expressions for both its mean in Eq.~(\ref{eq:Expectation}) and its second moment in Eq.~(\ref{second_moment}). 
These expressions depend only on the single-particle density and up to four-body correlation functions, yet fully capture the effects of strong coupling whenever the distribution of $V_{\mathcal{SE}}$ is near‑Gaussian. 
With these first moments of the distribution, we derived a closed‑form expression for the free‑energy shift between a strongly coupled system and its weakly coupled counterpart.
Importantly, Monte Carlo benchmarks across different system sizes confirmed that our model is able to span the entire coupling spectrum: from the delta‑function limit in large systems to the broad, fluctuation‑dominated behavior in small systems, with excellent agreement between our model and explicit simulation results. Direct access to $\langle V_{\mathcal{SE}}\rangle$ and $\mathrm{Var}(V_{\mathcal{SE}})$ therefore opens a practical path toward computing free energies, excess chemical potentials, and studying strongly coupled systems (see the companion article \cite{PRL}). 
Since the framework assumes only pairwise interactions, it can be extended to include internal degrees of freedom --- for example by embedding a quantum mechanical region within a classical environment (QM/MM)  \cite{csizi2023universal, tzeliou2022review} ---  
opening promising avenues for multiscale simulations that bridge classical and quantum descriptions. 
We propose extending our moment‑based reference‑particle framework to Quasichemical Theory \cite{rogers2008modeling,gomez2022hydrated}, in which the environment is partitioned into an inner shell (directly interacting particles) and an outer shell (bulk). 
This modular approach will provide a transparent and efficient pathway for computing solvation and binding free energies in complex fluids \cite{asthagiri2025consequences}.
\vspace{1cm}
\begin{acknowledgments}
We gratefully acknowledge ﬁnancial support by the DFG
under Germany’s Excellence Strategy EXC 2089/1-390776260
(e-conversion).
\end{acknowledgments}

\appendix
\section{Fundamentals of $n$-Particle Densities and Correlation Functions}\label{$n$-Particle Densities}

For clarity and completeness, we briefly recall the standard definitions of reduced density functions and correlation functions used throughout this work \cite{burgot2017notion, hansen2013theory, zwanzig2001nonequilibrium}. 
These quantities provide the link between the full microscopic description of an $N$-particle system and the coarse-grained statistical measures used in the main derivations. 
Let $P^{(N)}(q_1, q_2, \dots, q_N)$ denote the full $N$-particle probability density. 
This function gives the probability of observing particles in the infinitesimal volume element $\mathrm{d}^3q_1 \cdots \mathrm{d}^3q_N$ centered at positions $q_1, \dots, q_N$. 
In many-body systems, observables rarely depend on all particle coordinates simultaneously.
To isolate the statistics of a subset of $n<N$ particles, one defines the $n$-particle reduced density $P^{(n)}(q_1, \dots, q_n)$ by integrating out the remaining variables
\begin{equation}
P^{(n)}(q_1, \dots, q_n) = \int P^{(N)}(q_1, \dots, q_N)\, \mathrm{d}^3q_{n+1} \cdots \mathrm{d}^3q_N.
\end{equation}
This quantity refers to a particular group of $n$ labeled particles. 
For indistinguishable systems, it is often more natural to consider the generic $n$-particle density $\rho^{(n)}(q_1, \dots, q_n)$, defined as
\begin{equation}
\rho^{(n)}(q_1, \dots, q_n) = \frac{N!}{(N-n)!}\, P^{(n)}(q_1, \dots, q_n),
\end{equation}
which accounts for all unordered selections of $n$ particles located at the specified positions. 
To quantify correlations between particle positions, it is common to introduce the $n$-body correlation function
\begin{equation}
g^{(n)}(q_1, \dots, q_n) = \frac{\rho^{(n)}(q_1, \dots, q_n)}{\prod_{i=1}^n \rho^{(1)}(q_i)},
\end{equation}
where $\rho^{(1)}(q)$ denotes the single-particle density. 
The function $g^{(n)}$ measures deviations from a product distribution and encodes how spatial configurations differ from those of statistically independent particles. 
These definitions provide the structural backbone for interpreting the reduced expressions that appear in the main derivations, including pair and higher-order interactions between system and environment components.

\section{ Reference Particle Approach}\label{Arpa}
As discussed in the main text, the quantity $\langle \mathcal{A}_{\mathcal{S}\mathcal{E}} \rangle $ can be expressed using the reference-particle framework as
\begin{align}
\langle \mathcal{A}_{\mathcal{S}\mathcal{E}} \rangle 
&= \frac{1}{\Omega} \int_{V_\mathcal{S}} \mathrm{d} q_\odot \sum_{k=1}^N \mathcal{IL}^{N} \Big\{ \delta \mathcal{C}(x_\mathcal{E}, x_\mathcal{S}; E) \, \delta(q_\odot - q_k) \nonumber\\
&\qquad \times\, \mathcal{A'}_{\mathcal{S}\mathcal{E}}(q_k, \circledcirc) \Big\}.
\label{master}
\end{align}
This representation is built around the idea of systematically placing a reference particle at each position $q_\odot$ within the system volume $V_\mathcal{S}$, and evaluating the contribution $\mathcal{A'}_{\mathcal{S}\mathcal{E}}(q_k, \circledcirc)$. 
The term $\mathcal{A'}_{\mathcal{S}\mathcal{E}}$ captures the effect of the surrounding configuration, conditioned on particle $k$ being pinned at $q_\odot$, and the average is computed over all configurations consistent with that condition. 
This strategy transforms the global averaging problem into a sum of localized problems, each defined relative to a single tagged particle. 
In the special case where $\mathcal{A}_{\mathcal{S}\mathcal{E}}$ corresponds to an interaction energy, the function $\mathcal{A'}_{\mathcal{S}\mathcal{E}}$ encodes the interaction of a single particle with its environment. 
This localized viewpoint not only simplifies the computational structure, but also sets the stage for leveraging statistical quantities such as single-particle densities and many-body correlation functions --- tools that are essential for capturing how local particle interactions shape emergent macroscopic behavior. 
To formalize the above ideas, we introduce the generic single-particle probability density $\rho(q_\odot)$, defined by
\begin{equation}
\rho(q_\odot) = \frac{\sum_{k=1}^N \delta_{q_\odot q_k} \otimes \mathcal{L}_{q_k}^{N} \left\{ \delta \mathcal{C}(x_\mathcal{E}, x_\mathcal{S}; E) \right\}}{\Omega}.
\label{eq:density}
\end{equation}
This quantity represents the probability density of finding any particle of the system at position $q_\odot$, independent of its label. 
Next, we define the generic single-particle conditional expectation of $\mathcal{A}_{\mathcal{S}\mathcal{E}}$, given that a particle is fixed at $q_\odot$, as
\begin{align}
&\nonumber\left\langle \mathcal{A'}_{\mathcal{S}\mathcal{E}}(q_\odot , \circledcirc) \right\rangle_{\circledcirc | q_\odot} 
\\&= \frac{\sum_{k=1}^N \delta_{q_\odot q_k} \otimes \mathcal{L}_{q_k}^{N} \left\{ \delta \mathcal{C}(x_\mathcal{E}, x_\mathcal{S}; E) \, \mathcal{A'}_{\mathcal{S}\mathcal{E}}(q_k, \circledcirc) \right\}}{\sum_{k=1}^N \delta_{q_\odot q_k} \otimes \mathcal{L}_{q_k}^{N} \left\{ \delta \mathcal{C}(x_\mathcal{E}, x_\mathcal{S}; E) \right\}}.
\label{eq:cond_exp}
\end{align}
Substituting Eqs.~(\ref{eq:density}) and~(\ref{eq:cond_exp}) into the global expectation formula in Eq.~(\ref{master}), we obtain
\begin{equation}
\left\langle \mathcal{A}_{\mathcal{S}\mathcal{E}} \right\rangle 
= \int_{V_\mathcal{S}} \mathrm{d}q_\odot \, \rho(q_\odot) \left\langle \mathcal{A'}_{\mathcal{S}\mathcal{E}}(q_\odot, \circledcirc) \right\rangle_{\circledcirc | q_\odot}.
\end{equation}
This result shows that the total expectation is a weighted average of localized contributions from each possible reference position $q_\odot$, with weights given by the generic single-particle probability density $\rho(q_\odot)$. 
To simplify the generic conditional expectation $\langle \mathcal{A'}_{\mathcal{S}\mathcal{E}}(q_\odot, \circledcirc) \rangle_{\circledcirc|q_\odot}$, we first rewrite it by substituting the definition of $\rho(q_\odot)$ from Eq.~(\ref{eq:density}) into the denominator of Eq.~(\ref{eq:cond_exp}). 
This yields
\begin{align}
\langle \mathcal{A'}_{\mathcal{S}\mathcal{E}}(q_\odot, \circledcirc) \rangle_{\circledcirc|q_\odot} &\nonumber=\sum_{k=1}^N \mathcal{L}_{q_k}^{N} \{ \frac{\delta_{q_\odot q_k} \otimes\delta \mathcal{C}(x_\mathcal{E}, x_\mathcal{S}; E)}{\rho(q_\odot)\Omega} \\& \times\delta_{q_\odot q_k} \otimes\mathcal{A'}_{\mathcal{S}\mathcal{E}}({q_k},\circledcirc) \}. \label{eq:cond_exp_with_density}
\end{align}
Recognizing that ${\delta \mathcal{C}(x_\mathcal{E}, x_\mathcal{S}; E)}/{\Omega}$ defines the probability distribution $P(x_\mathcal{E}, x_\mathcal{S})$ over microstates of the composite system in the microcanonical ensemble, we rewrite Eq.~(\ref{eq:cond_exp_with_density}) as
\begin{align}
\langle \mathcal{A'}_{\mathcal{S}\mathcal{E}}(q_\odot, \circledcirc) \rangle_{\circledcirc|q_\odot} &\nonumber= \sum_{k=1}^N \mathcal{L}_{q_k}^{N} \{ \frac{\delta_{q_\odot q_k} \otimes P(x_\mathcal{E}, x_\mathcal{S})}{\rho(q_\odot)} \\&\times \delta_{q_\odot q_k} \otimes\mathcal{A'}_{\mathcal{S}\mathcal{E}}({q_k},\circledcirc) \}. \label{eq:joint_prob_sub}
\end{align}
Next, we apply Bayes’ theorem to factorize $P(x_\mathcal{E}, x_\mathcal{S})$
\begin{equation}
P(x_\mathcal{E}, x_\mathcal{S}) = P(q_k) P(\circledcirc | q_k), \label{eq:bayes_factorization}
\end{equation}
where $P(q_k)$ is the specific single-particle probability density which can be written as
\begin{equation}
P(q_k)= \frac{\mathcal{L}_{q_k}^{N} \{\delta \mathcal{C}(x_\mathcal{E}, x_\mathcal{S}; E)\}}{\Omega}
\label{specific_probability}
\end{equation}
By comparing Eq.~(\ref{specific_probability}) with Eq.~(\ref{eq:density}), one finds that
\begin{equation}
\rho(q_\odot) = \sum_{k=1}^N \delta_{q_\odot q_k} \otimes P(q_k). \label{eq:spatial_density_def}
\end{equation}
Substituting Eqs.~(\ref{eq:bayes_factorization}) and (\ref{eq:spatial_density_def}) into Eq.~(\ref{eq:joint_prob_sub}) gives
\begin{align}
\langle \mathcal{A'}_{\mathcal{S}\mathcal{E}}(q_\odot, \circledcirc) \rangle_{\circledcirc|q_\odot} &\nonumber= \sum_{k=1}^N \mathcal{L}_{q_k}^{N} \{ \frac{\delta_{q_\odot q_k} \otimes P(q_k)}{\sum_{k=1}^N \delta_{q_\odot q_k} \otimes P(q_k)} \\&\nonumber\times\delta_{q_\odot q_k} \otimes 
P(\circledcirc | q_k)\\&\times \delta_{q_\odot q_k} \otimes \mathcal{A'}_{\mathcal{S}\mathcal{E}}({q_k},\circledcirc) \}. \label{eq:cond_prob_final}
\end{align}   
Since the particles are identical, the contribution from any particle at $q_\odot$ is uniform. That leads to 
\begin{equation}
\frac{\delta_{q_\odot q_k} \otimes P(q_k)}{\sum_{k=1}^N \delta_{q_\odot q_k} \otimes P(q_k)} = \frac{1}{N}.
\label{uniform}
\end{equation}
Substituting Eq.~(\ref{uniform}) into Eq.~(\ref{eq:cond_prob_final}), we obtain
\begin{align}
&\nonumber\langle \mathcal{A'}_{\mathcal{S}\mathcal{E}}(q_\odot, \circledcirc) \rangle_{\circledcirc|q_\odot}\\[6pt]
&= \frac{1}{N}\sum_{k=1}^N \delta_{q_\odot q_k} \otimes \mathcal{L}_{q_k}^{N} \left\{ P(\circledcirc | q_k) \mathcal{A'}_{\mathcal{S}\mathcal{E}}({q_k},\circledcirc) \right\} . \label{eq:cond_exp_simplified}
\end{align}
The expression can also be written in a more compact form by introducing the specific single-particle conditional expectation
\begin{equation}
\langle \mathcal{A'}_{\mathcal{S}\mathcal{E}}({q_k}, \circledcirc) \rangle_{\circledcirc \mid q_k} := 
\mathcal{L}_{q_k}^{N} \left\{ P(\circledcirc \mid q_k)\, \mathcal{A'}_{\mathcal{S}\mathcal{E}}({q_k}, \circledcirc) \right\}.
\end{equation}
Substituting this into Eq.~(\ref{eq:cond_exp_simplified}) yields
\begin{align}
\langle \mathcal{A'}_{\mathcal{S}\mathcal{E}}(q_\odot, \circledcirc) \rangle_{\circledcirc \mid q_\odot}
= \frac{1}{N} \sum_{k=1}^N \delta_{q_\odot q_k} \otimes 
\langle \mathcal{A'}_{\mathcal{S}\mathcal{E}}({q_k}, \circledcirc) \rangle_{\circledcirc \mid q_k}.
\label{eq:final_cond_exp}
\end{align}
Due to the indistinguishability of particles, each term in the sum contributes equally and we obtain
\begin{equation}
\langle \mathcal{A'}_{\mathcal{S}\mathcal{E}}(q_\odot, \circledcirc) \rangle_{\circledcirc \mid q_\odot} 
= \delta_{q_\odot q_k} \otimes 
\langle \mathcal{A'}_{\mathcal{S}\mathcal{E}}({q_k}, \circledcirc) \rangle_{\circledcirc \mid q_k}.
\label{eq:final_result}
\end{equation}
This final identity connects the generic single-particle conditional expectation to the specific one, highlighting that --- due to particle indistinguishability --- the local contributions can be reduced to a single tagged-particle average evaluated at $q_\odot$. Consequently, the full expectation of the observable takes the form
\begin{align}
\langle \mathcal{A}_{\mathcal{S}\mathcal{E}} \rangle 
= \int_{V_\mathcal{S}} \mathrm{d}q_\odot \, \rho(q_\odot)\,
\delta_{q_\odot q_k} \otimes 
\langle \mathcal{A'}_{\mathcal{S}\mathcal{E}}(q_k, \circledcirc) \rangle_{\circledcirc \mid q_k}.
\end{align}

\section{Derivation of the First Moment of Interaction}\label{Derivation of the first moment of Interaction}
In this section, we provide a detailed derivation of Eq.~(\ref{eq:Expectation}), starting from the general formulation in Eq.~(\ref{V_expect}). 
Using the reference-particle framework, the average interaction energy can be expressed as
\begin{equation}
\langle V_{\mathcal{S}\mathcal{E}} \rangle = \int_{V_\mathcal{S}} \mathrm{d}q_\odot \, \rho(q_\odot)\, \delta_{q_\odot q_k} \otimes \langle \mathcal{A'}_{\mathcal{S}\mathcal{E}}(q_k, \circledcirc) \rangle_{\circledcirc \mid q_k},
\label{total V}
\end{equation}
where the conditional expectation is given as (see Eq.~(\ref{A_exp}))
\begin{align}
\langle \mathcal{A'}_{\mathcal{S}\mathcal{E}}(q_k, \circledcirc) \rangle_{\circledcirc | q_k}
&\nonumber= \int_{V_\mathcal{E}} \mathrm{d}q \sum_{\substack{j=1 \\ j \neq k}}^{N} 
\mathcal{L}_{q_k}^{\circledcirc} \left\{ P(\circledcirc \mid q_k)\, \delta(q - q_j) \right\}
\\&\times u\left( \| q - q_k \|_2 \right).
\label{A_ave}
\end{align}
Using the property defined in Eq.~(\ref{eq:reduced_cond_expectation}), we simplify the inner term
\begin{align}
&\nonumber \sum_{\substack{j=1\\ j\neq k}}^{N} \mathcal{L}_{q_k}^{N} \left\{ P(\circledcirc | q_k)\delta(q - q_j) \right\} 
= \sum_{\substack{j=1\\ j\neq k}}^{N} \int \mathrm{d}q_j \, \delta(q - q_j) P(q_j | q_k) 
\\&= \sum_{\substack{j=1\\ j\neq k}}^{N} \frac{P(q, q_k)}{P(q_k)} = (N-1)\frac{P(q, q_k)}{P(q_k)}.
\end{align}
The final expression can be rewritten using the relation between specific and generic reduced distributions (see Appendix~\ref{$n$-Particle Densities})
\begin{equation}
(N-1)\frac{P(q, q_k)}{P(q_k)} = \frac{\rho(q, q_k)}{\rho(q_k)} = \rho(q)\, g^{(2)}(q, q_k).
\end{equation}
Substituting this into Eq.~(\ref{A_ave}), we obtain
\begin{equation}
\langle \mathcal{A'}_{\mathcal{S}\mathcal{E}}(q_k, \circledcirc) \rangle_{\circledcirc \mid q_k}
= \int_{V_\mathcal{E}} \mathrm{d}q\, \rho(q)\, g^{(2)}(q, q_k)\, u(\| q - q_k \|_2).
\end{equation}
Finally, inserting this result into Eq.~(\ref{total V}), we arrive at
\begin{align}
\langle V_{\mathcal{S}\mathcal{E}} \rangle 
&= \int_{V_\mathcal{S}} \mathrm{d}q_\odot \, \rho(q_\odot) \int_{V_\mathcal{E}} \mathrm{d}q \, \rho(q)\, g^{(2)}(q, q_\odot)\, u(\|q - q_\odot \|_2),
\end{align}
which is the final result of Eq.~(\ref{eq:Expectation}) in the main text.
\section{Derivation of the Second Moment of Interaction }\label{Derivation of the Second Moment of Interaction}

We begin by examining the diagonal contribution $ \mathcal{D}_{\mathcal{S}\mathcal{E}}(q_k, \circledcirc) $, which, by definition, accounts for the squared interaction of the reference particle $ q_k $ with the environment. 
This term, as introduced in Eq.~(\ref{D}), can be written explicitly as
\begin{align}
\mathcal{D}_{\mathcal{S}\mathcal{E}}(q_k, \circledcirc ) &= \int_{V_\mathcal{E}} \mathrm{d}q \int_{V_\mathcal{E}} \mathrm{d}q' \sum_{\substack{j=1 \\ j \neq k}}^{N} \sum_{\substack{j'=1 \\ j' \neq k}}^{N} \delta(q - q_j)\, \delta(q' - q_{j'}) \nonumber \\
&\quad \times u\left( \lVert q - q_k \rVert_2 \right)\, u\left( \lVert q' - q_k \rVert_2 \right).
\end{align}
The double summation over indices $ j $ and $ j' $ enumerates all possible pairs of environment particles that interact with the same reference particle $ q_k $. To isolate structurally distinct terms, we rewrite this sum as
\begin{equation}
\sum_{\substack{j=1\\ j\neq k}}^{N} \sum_{\substack{j'=1\\ j'\neq k}}^{N} 
= \sum_{\substack{j=1\\ j\neq k}}^{N} \sum_{\substack{j'=1\\ j'\neq k \\ j'\neq j}}^{N} 
+ \sum_{\substack{j=1\\ j\neq k \\ j'=j}}^{N}.
\end{equation} 
The first term collects cross-pairings with different environment particles, while the second isolates the interaction with the same particle  $ j = j' $, both of which will enter the evaluation of $ \langle \mathcal{D}_{\mathcal{S}\mathcal{E}}(q_k, \circledcirc) \rangle_{\circledcirc | q_k} $ below. Applying the specific single-particle conditional average $ \langle \cdots \rangle_{\circledcirc|q_k} $, we obtain
\begin{align}
\langle\mathcal{D}_{\mathcal{S}\mathcal{E}}(q_k, \circledcirc ) \rangle_{\circledcirc|q_k} &\nonumber= \int_{V_\mathcal{E}} \mathrm{d}q \int_{V_\mathcal{E}} \mathrm{d}q' \sum_{\substack{j=1\\ j\neq k}}^{N} \sum_{\substack{j'=1\\ j'\neq k\\j'\neq j} }^{N}\\&\nonumber\times\mathcal{L}_{q_k}^{N} \{P(\circledcirc|q_k)\delta(q - q_j) \delta(q' - q_{j'})\} \\&\nonumber\times u\left(\lVert q - q_k \rVert_2 \right) u\left(\lVert q' - q_k \rVert_2 \right)\\&\nonumber +  \int_{V_\mathcal{E}} \mathrm{d}q \int_{V_\mathcal{E}} \mathrm{d}q' \sum_{\substack{j=1\\ j\neq k}}^{N} \\&\nonumber\times\mathcal{L}_{q_k}^{N} \{P(\circledcirc|q_k)\delta(q - q_j) \delta(q' - q_j)\} \\&\times u\left(\lVert q - q_k \rVert_2 \right) u\left(\lVert q' - q_k \rVert_2 \right).
\label{D_main}
\end{align}
To simplify the above expression, we follow the same idea employed in Eq.~(\ref{A_ave}).
Specifically, we make use of the property introduced in Eq.~(\ref{eq:reduced_cond_expectation}), along with the relation between specific and generic reduced distributions (see Appendix \ref{$n$-Particle Densities} for details).
To avoid redundancies, we ask the reader to keep this reasoning in mind when analyzing the remaining expressions in this Appendix. 
For distinct indices $ j $ and $ j' $, we apply the above logic and obtain
\begin{align}
&\nonumber\sum_{\substack{j=1\\ j\neq k}}^{N} \sum_{\substack{j'=1\\ j'\neq k\\j'\neq j} }^{N} \mathcal{L}_{q_k}^{N} \{P(\circledcirc|q_k)\delta(q - q_j) \delta(q' - q_{j'})\} \\&\nonumber=  \sum_{\substack{j=1\\ j\neq k}}^{N} \sum_{\substack{j'=1\\ j'\neq k\\j'\neq j} }^{N}  \int\int \mathrm{d}q_j \mathrm{d}q_{j'}\delta(q - q_j) \delta(q' - q_{j'}) P(q_j, q_{j'}|q_k) \\&\nonumber= (N-1)(N-2)\frac{P(q,q',q_k)}{P(q_k)} =\frac{\frac{N!}{(N-3)!}P(q,q',q_k)}{\frac{N!}{(N-1)!}P(q_k)}\\&=\frac{\rho(q,q',q_k)}{\rho(q_k)} = \rho(q)\rho(q')g^{(3)}(q,q', q_k).
\label{D_1}
\end{align}
Similarly, for $j'=j$, we can write
\begin{align}
&\nonumber\sum_{\substack{j=1\\ j\neq k}}^{N} \mathcal{L}_{q_k}^{N} \{P(\circledcirc|q_k)\delta(q - q_j) \delta(q' - q_j)\} \\&=\delta(q - q') \rho(q')g^{(2)}(q',q_k).
\label{D_2}
\end{align}
Substituting Eq.~(\ref{D_1}) and Eq.~(\ref{D_2}) into Eq.~(\ref{D_main}), and reinserting the result into the diagonal integral using Eq.~(\ref{central}) and Eq.~(\ref{D}), we obtain
\begin{align}
\mathcal{D} &\nonumber=\int_{V_\mathcal{S}} \mathrm{d}q_\odot \, \rho(q_\odot) \int_{V_\mathcal{E}}\mathrm{d}q \int_{V_\mathcal{E}}\mathrm{d}q' \rho(q)\rho(q')\\&\times g^{(3)}(q,q', q_\odot) \nonumber u(\|q - q_\odot\|_2)u(\|q' - q_\odot\|_2) \\&\nonumber+ \int_{V_\mathcal{S}} \mathrm{d}q_\odot \, \rho(q_\odot) \int_{V_\mathcal{E}} \mathrm{d}q \,\rho(q)\\&\times g^{(2)}(q, q_\odot)u(\|q - q_\odot\|_2)^2.
\end{align}
which corresponds to the sum of the two contributions $M$ and $R$ defined earlier in Eq.~(\ref{eq:M_term}) and Eq.~(\ref{eq:R_term}). 
We now turn our attention to the off-diagonal contribution $\mathcal{I}_{\mathcal{S}\mathcal{E}}(q_k, \circledcirc)$ appearing in the decomposition of $V_{\mathcal{S}\mathcal{E}}^2$.
\begin{align}
\mathcal{I}_{\mathcal{S}\mathcal{E}}(q_k, \circledcirc) 
&\nonumber= \int_{V_\mathcal{S}} \mathrm{d}q_* \int_{V_\mathcal{E}} \mathrm{d}q \int_{V_\mathcal{E}} \mathrm{d}q' \sum_{\substack{j=1\\ j\neq k}}^{N} 
\sum_{\substack{j'=1\\ j'\neq k}}^{N} \sum_{\substack{j''=1\\ j''\neq k}}^{N} \\
&\nonumber\quad \times \delta(q - q_j) \delta(q' - q_{j'}) \delta(q_* - q_{j''}) \\
&\quad \times u\left(\lVert q - q_k \rVert_2 \right) u\left(\lVert q' - q_* \rVert_2 \right).
\end{align}
The triple summation over $j$, $j'$, and $j''$ spans all possible environment particles not equal to $k$. 
However, it is important to mention that the presence of multiple indices introduces distinct combinatorial scenarios.
To systematically capture these cases, we decompose the triple sum into five disjoint classes, each corresponding to a unique combination of index constraints
\begin{align}
&\nonumber\sum_{\substack{j=1\\ j\neq k}}^{N} \sum_{\substack{j'=1\\ j'\neq k}}^{N} \sum_{\substack{j''=1\\ j''\neq k}}^{N}  = \sum_{\substack{j=1\\ j\neq k}}^{N} \sum_{\substack{j'=1\\ j'\neq k \\ j'\neq j}}^{N} \sum_{\substack{j''=1\\ j''\neq k\\ j''\neq j'}}^{N} + \sum_{\substack{j=1\\ j\neq k \\ j'=j\\j''=j}}^{N} \\&+\sum_{\substack{j=1\\ j\neq k \\j''=j }}^{N} \sum_{\substack{j'=1\\ j'\neq k\\j'\neq j}}^{N} + \sum_{\substack{j=1\\ j\neq k\\j'=j }}^{N} \sum_{\substack{j''=1\\ j''\neq k\\j''\neq j}}^{N} + \sum_{\substack{j''=1\\ j''\neq k\\j' = j'' }}^{N} \sum_{\substack{j=1\\ j\neq k\\j\neq j''}}^{N}.
\end{align}
Now, we apply the specific single-particle conditional average $\langle \cdots \rangle_{\circledcirc|q_k}$ and obtain
\begin{widetext}
\begin{align}
&\nonumber\langle\mathcal{I}_{\mathcal{S}\mathcal{E}}(q_k, \circledcirc) \rangle_{\circledcirc|q_k} \\&\nonumber=    \int_{V_\mathcal{S}} \mathrm{d}q_* \int_{V_\mathcal{E}} \mathrm{d}q \int_{V_\mathcal{E}} \mathrm{d}q'   {\sum_{\substack{j=1\\ j\neq k}}^{N} \sum_{\substack{j'=1\\ j'\neq k \\ j'\neq j}}^{N} \sum_{\substack{j''=1\\ j''\neq k\\ j''\neq j'}}^{N}  \mathcal{L}_{q_k}^{N} \{P(\circledcirc|q_k)\delta(q - q_j) \delta(q' - q_{j'}) \delta(q_* - q_{j''}) } \}u\left(\lVert q - q_k \rVert_2 \right) u\left(\lVert q' - q_* \rVert_2 \right)\\&\nonumber +  \int_{V_\mathcal{S}} \mathrm{d}q_* \int_{V_\mathcal{E}} \mathrm{d}q \int_{V_\mathcal{E}} \mathrm{d}q'   {\sum_{\substack{j=1\\ j\neq k }}^{N}\mathcal{L}_{q_k}^{N} \{P(\circledcirc|q_k)   \delta(q - q_j) \delta(q' - q_{j}) \delta(q_* - q_{j})} \}u(\| q - q_k \|_2 ) u(\| q' - q_*\|_2) \\&\nonumber+   \int_{V_\mathcal{S}} \mathrm{d}q_* \int_{V_\mathcal{E}} \mathrm{d}q \int_{V_\mathcal{E}} \mathrm{d}q'   {\sum_{\substack{j=1\\ j\neq k }}^{N} \sum_{\substack{j'=1\\ j'\neq k\\j'\neq j}}^{N}   \mathcal{L}_{q_k}^{N} \{P(\circledcirc|q_k)\delta(q - q_j) \delta(q' - q_{j'}) \delta(q_* - q_{j}) } \}u(\| q - q_k \|_2 ) u(\| q' - q_*\|_2)\\&\nonumber+   \int_{V_\mathcal{S}} \mathrm{d}q_* \int_{V_\mathcal{E}} \mathrm{d}q \int_{V_\mathcal{E}} \mathrm{d}q'   {\sum_{\substack{j=1\\ j\neq k }}^{N} \sum_{\substack{j''=1\\ j''\neq k\\j''\neq j}}^{N}  \mathcal{L}_{q_k}^{N} \{P(\circledcirc|q_k) \delta(q - q_j) \delta(q' - q_{j}) \delta(q_* - q_{j''}) \} }u(\| q - q_k \|_2) u(\|q' - q_*\|_2)\\&+   \int_{V_\mathcal{S}} \mathrm{d}q_* \int_{V_\mathcal{E}} \mathrm{d}q \int_{V_\mathcal{E}} \mathrm{d}q'   {\sum_{\substack{j''=1\\ j''\neq k }}^{N} \sum_{\substack{j=1\\ j\neq k\\j\neq j''}}^{N}  \mathcal{L}_{q_k}^{N} \{P(\circledcirc|q_k) \delta(q - q_{j}) \delta(q' - q_{j''}) \delta(q_* - q_{j''})\}} u(\|q - q_k\|_2) u(\|q' - q_*\|_2).
\end{align}
\end{widetext}

As before, we systematically evaluate each term by applying the transformation rules from Eq.~(\ref{eq:reduced_cond_expectation}) and using the relation between specific and generic reduced distributions. We begin with the term corresponding to the case in which all three indices $j$, $j'$, and $j''$ are distinct. 
\begin{widetext}
\begin{align}
&\nonumber{\sum_{\substack{j=1\\ j\neq k}}^{N} \sum_{\substack{j'=1\\ j'\neq k \\ j'\neq j}}^{N} \sum_{\substack{j''=1\\ j''\neq k\\ j''\neq j'}}^{N}  \mathcal{L}_{q_k}^{N} \{P(\circledcirc|q_k)\delta(q - q_j) \delta(q' - q_{j'}) \delta(q_* - q_{j''}) } \} \\&\nonumber=  \sum_{\substack{j=1\\ j\neq k}}^{N} \sum_{\substack{j'=1\\ j'\neq k \\ j'\neq j}}^{N} \sum_{\substack{j''=1\\ j''\neq k\\ j''\neq j'}}^{N} \int\int\int \mathrm{d}q_j \mathrm{d}q_{j'}\mathrm{d}q_{j''}\delta(q - q_j) \delta(q^\prime - q_{j'}) \delta(q_* - q_{j''})  P(q_j, q_{j'}, q_{j''} |q_k)\\&\nonumber=\sum_{\substack{j=1\\ j\neq k}}^{N} \sum_{\substack{j'=1\\ j'\neq k \\ j'\neq j}}^{N} \sum_{\substack{j''=1\\ j''\neq k\\ j''\neq j'}}^{N} \frac{1}{P(q_k)} \int\int\int \mathrm{d}q_j \mathrm{d}q_{j'} \mathrm{d}q_{j''}\delta(q - q_j) \delta(q^\prime - q_{j'}) \delta(q_* - q_{j''})  P(q_j, q_{j'}, q_{j''}, q_k)\\& \nonumber= (N-1) (N-2) (N-3 ) \frac{P(q, q', q_*, q_k)}{P(q_k)} = \frac{\frac{N!}{(N-4)!}\,P(q, q', q_*, q_k)}{\frac{N!}{(N-1)!}P(q_k)} \\&= \frac{\rho(q, q', q_*, q_k)}{\rho(q_k)} = \rho(q)\rho(q')\rho(q_*) g^{(4)}(q,q',q_*,q_k)
\end{align}
\end{widetext}
We then proceed to the contribution where all three indices coincide ($j'' = j' = j$). 
\begin{widetext}
\begin{align}
&\nonumber\sum_{\substack{j=1\\ j\neq k }}^{N}\mathcal{L}_{q_k}^{N} \{P(\circledcirc|q_k)   \delta(q - q_j) \delta(q' - q_{j}) \delta(q_* - q_{j}) \\&\nonumber= \sum_{\substack{j=1\\ j\neq k }}^{N}  \int \mathrm{d}q_j \delta(q - q_j) \delta(q^\prime - q_{j}) \delta(q_* - q_{j}) P(q_j|q_k) \\&\nonumber =\sum_{\substack{j=1\\ j\neq k }}^{N}   \frac{1}{P(q_k)} \int \mathrm{d}q_j  \delta(q - q_j) \delta(q^\prime - q_{j}) \delta(q_* - q_{j}) P(q_j, q_k)  \\&\nonumber= (N-1) \delta(q - q_*) \delta(q^\prime - q_*) \frac{P(q_*, q_k)}{P(q_k)} =\delta(q - q_*) \delta(q^\prime - q_*) \frac{\frac{N!}{(N-2)!}\,P(q_*, q_k)}{\frac{N!}{(N-1)!}P(q_k)}  \\&= \delta(q - q_*) \delta(q^\prime - q_*) \frac{\rho(q_*, q_k)}{\rho(q_k)}=  \delta(q - q_*) \delta(q^\prime - q_*) \rho (q_*)g^{(2)}(q_*,q_k) 
\end{align}
\end{widetext}
Next, we consider the first of the three terms, in which two of the indices coincide, specifically $j'' = j$ while $j' \neq j$.
\begin{widetext}
\begin{align}
&\nonumber\sum_{\substack{j=1\\ j\neq k }}^{N} \sum_{\substack{j'=1\\ j'\neq k\\j'\neq j}}^{N}   \mathcal{L}_{q_k}^{N} \{P(\circledcirc|q_k)\delta(q - q_j) \delta(q' - q_{j'}) \delta(q_* - q_{j}) \\&\nonumber= \sum_{\substack{j=1\\ j\neq k }}^{N} \sum_{\substack{j'=1\\ j'\neq k\\j'\neq j}}^{N} \int \int \mathrm{d}q_j \mathrm{d}q_{j'}\delta(q - q_j) \delta(q^\prime - q_{j'}) \delta(q_* - q_{j}) P(q_j, q_j' |q_k)\\&\nonumber = \sum_{\substack{j=1\\ j\neq k }}^{N} \sum_{\substack{j'=1\\ j'\neq k\\j'\neq j}}^{N} \frac{1}{P(q_k)} \int \int \mathrm{d}q_j \mathrm{d}q_{j'} \delta(q - q_j) \delta(q^\prime - q_{j'}) \delta(q_* - q_{j})  P(q_j, q_{j'}, q_k)  \\&\nonumber= (N-1) (N-2) \delta(q - q_*) \frac{P(q_*, q', q_k)}{P(q_k)} =\delta(q - q_*) \frac{\frac{N!}{(N-3)!}\,P(q_*, q', q_k)}{\frac{N!}{(N-1)!}P(q_k)}  \\&= \delta(q - q_*) \frac{\rho(q_*, q', q_k)}{\rho(q_k)} =  \delta(q - q_*) \rho (q_*)\rho (q')g^{(3)}(q_*, q', q_k) 
\end{align}
\end{widetext}
 For the case where $j' = j$ and $j'' \neq j$ we obtain
\begin{widetext}
\begin{align}
&\nonumber\sum_{\substack{j=1\\ j\neq k }}^{N} \sum_{\substack{j''=1\\ j''\neq k\\j''\neq j}}^{N}  \mathcal{L}_{q_k}^{N} \{P(\circledcirc|q_k) \delta(q - q_j) \delta(q' - q_{j}) \delta(q_* - q_{j''}) \\&\nonumber=\sum_{\substack{j=1\\ j\neq k }}^{N} \sum_{\substack{j''=1\\ j''\neq k\\j''\neq j}}^{N}  \int \int \mathrm{d}q_j \mathrm{d}q_{j''}\delta(q - q_j) \delta(q^\prime - q_{j}) \delta(q_* - q_{j''}) P(q_j, q_{j''} |q_k)\\&\nonumber \\&\nonumber= \sum_{\substack{j=1\\ j\neq k }}^{N} \sum_{\substack{j''=1\\ j''\neq k\\j''\neq j}}^{N} \frac{1}{P(q_k)} \int \int \mathrm{d}q_j \mathrm{d}q_{j''} \delta(q - q_j) \delta(q^\prime - q_{j}) \delta(q_* - q_{j''}) P(q_j, q_{j''}, q_k)  \\&\nonumber = (N-1) (N-2) \delta(q - q') \frac{P(q', q_*, q_k)}{P(q_k)}  =\delta(q - q') \frac{\frac{N!}{(N-3)!}\,P(q', q_*, q_k)}{\frac{N!}{(N-1)!}P(q_k)} \\& = \delta(q - q') \frac{\rho(q', q_*, q_k)}{\rho (q_k)}= \delta(q - q')\rho (q') \rho (q_*)g^{(3)}(q', q_*, q_k),
\end{align}
\end{widetext}
and finally, for the case where $j' = j''$ while $j \neq j''$, we arrive at
\begin{widetext}
\begin{align}
&\nonumber\sum_{\substack{j''=1\\ j''\neq k }}^{N} \sum_{\substack{j=1\\ j\neq k\\j\neq j''}}^{N}  \mathcal{L}_{q_k}^{N} \{P(\circledcirc|q_k) \delta(q - q_{j}) \delta(q' - q_{j''}) \delta(q_* - q_{j''}) \\&\nonumber= \sum_{\substack{j''=1\\ j''\neq k }}^{N} \sum_{\substack{j=1\\ j\neq k\\j\neq j''}}^{N} \int \int \mathrm{d}q_{j''} \mathrm{d}q_{j}\delta(q - q_{j}) \delta(q^\prime - q_{j''}) \delta(q_* - q_{j''})P(q_{j}, q_j'' |q_k) \\&\nonumber = \sum_{\substack{j''=1\\ j''\neq k }}^{N} \sum_{\substack{j=1\\ j\neq k\\j\neq j''}}^{N}\frac{1}{P(q_k)} \int \int \mathrm{d}q_{j''} \mathrm{d}q_{j}   \delta(q - q_{j}) \delta(q^\prime - q_{j''}) \delta(q_* - q_{j''}) P(q_{j}, q_{j''}, q_k) \\&\nonumber = (N-1) (N-2) \delta(q' - q_*) \frac{P(q, q_*, q_k)}{P(q_k)}  =\delta(q' - q_*) \frac{\frac{N!}{(N-3)!}\,P(q, q_*, q_k)}{\frac{N!}{(N-1)!}P(q_k)} \\&= \delta(q' - q_*) \frac{\rho(q, q_*, q_k)}{\rho(q_k)} = \delta(q' - q_*) \rho (q)\rho (q_*)g^{(3)}(q, q_*, q_k) 
\end{align}
\end{widetext}
Substituting the results from each of the five distinct summation scenarios -- classified based on coinciding index -- into the conditional expectation, and reinserting it into the off-diagonal integral using Eq.~(\ref{central}) and Eq.~(\ref{I}), we obtain the complete expression for the off-diagonal term $\mathcal{I}$. 
 The resulting expression for $\mathcal{I}$ is
\begin{widetext}
\begin{align}
\mathcal{I}&\nonumber=\int_{V_\mathcal{S}} \mathrm{d}q_\odot \, \rho(q_\odot) \left[\int_{V_\mathcal{S}}\mathrm{d}q_* \int_{V_\mathcal{E}}\mathrm{d}q \int_{V_\mathcal{E}}\mathrm{d}{q^\prime
} \rho(q)\rho(q')\rho(q_*) g^{(4)}(q,q',q_*,q_\odot)  \right] u\left(\lVert q - q_\odot \rVert_2\right) u\left(\lVert q^\prime - q_*\rVert_2\right) \\&\nonumber + \int_{V_\mathcal{S}} \mathrm{d}q_\odot \, \rho(q_\odot)\left[\underbrace{\int_{V_\mathcal{S}}\mathrm{d}q_* \int_{V_\mathcal{E}}\mathrm{d}q \int_{V_\mathcal{E}}\mathrm{d}{q^\prime
} \delta(q - q_*) \delta(q^\prime - q_*) \rho (q_*)g^{(2)}(q_*,q_\odot)}_{0}\right]u\left(\lVert q - q_\odot \rVert_2\right) u\left(\lVert q^\prime - q_*\rVert_2\right) \\&\nonumber+ \int_{V_\mathcal{S}} \mathrm{d}q_\odot \, \rho(q_\odot)\left[\underbrace{\int_{V_\mathcal{S}}\mathrm{d}q_* \int_{V_\mathcal{E}}\mathrm{d}q \int_{V_\mathcal{E}}\mathrm{d}{q^\prime
} \delta(q - q_*) \rho (q_*)\rho (q')g^{(3)}(q_*, q', q_\odot) }_{0} \right] u\left(\lVert q - q_\odot \rVert_2\right) u\left(\lVert q^\prime - q_*\rVert_2\right)  \\&\nonumber + \int_{V_\mathcal{S}} \mathrm{d}q_\odot \, \rho(q_\odot) \left[ \int_{V_\mathcal{S}}\mathrm{d}q_* \int_{V_\mathcal{E}}\mathrm{d}q \int_{V_\mathcal{E}}\mathrm{d}{q^\prime
} \delta(q - q')\rho (q') \rho (q_*)g^{(3)}(q', q_*, q_\odot) \right] u\left(\lVert q - q_\odot \rVert_2\right) u\left(\lVert q^\prime - q_*\rVert_2\right)\\&+\int_{V_\mathcal{S}} \mathrm{d}q_\odot \, \rho(q_\odot)\left[\underbrace{\int_{V_\mathcal{S}}\mathrm{d}q_* \int_{V_\mathcal{E}}\mathrm{d}q \int_{V_\mathcal{E}}\mathrm{d}{q^\prime
} \delta(q' - q_*) \rho (q)\rho (q_*)g^{(3)}(q, q_*, q_\odot) }_{0} \right] u\left(\lVert q - q_\odot \rVert_2\right) u\left(\lVert q^\prime - q_*\rVert_2\right) 
\end{align}
\end{widetext}
In the second term, the presence of $\delta(q - q_*)$ and $\delta(q' - q_*)$ imposes the condition that $q = q' = q_*$. However, $q$ and $q'$ are integrated over the environment volume $V_\mathcal{E}$, while $q_*$ is integrated over the system volume $V_\mathcal{S}$. Since the $V_\mathcal{E}$ and $V_\mathcal{S}$ are spatially disjoint, this condition cannot be satisfied within the integration domains, and the integral therefore vanishes. A similar reasoning eliminates the third and fifth terms as well. 
Hence, among the five possible index configurations, only two yield non-vanishing contributions due to the spatial disjointness of the system and environment volumes.
Discarding these zero contributions, the remaining terms capture the essential correlations that contribute to the fluctuation of the interaction energy between the system and the environment.
\begin{widetext}
\begin{align}
\mathcal{I}&\nonumber= \int_{V_\mathcal{S}} \mathrm{d}q_\odot \, \rho(q_\odot) \int_{V_\mathcal{S}} \mathrm{d}q_*\, \rho(q_*) \int_{V_\mathcal{E}} \, \mathrm{d}q' \rho(q') \int_{V_\mathcal{E}} \mathrm{d}q \,\rho(q) g^{(4)}(q_\odot, q, q^\prime , q_*) u\left(\lVert q - q_\odot\rVert_2\right) u\left(\lVert q^\prime - q_*\rVert_2\right)  \\& + \int_{V_\mathcal{S}} \mathrm{d}q_\odot \, \rho(q_\odot) \int_{V_\mathcal{S}} \mathrm{d}q_*\, \rho(q_*)  \int_{V_\mathcal{E}} \mathrm{d}q \,\rho(q) g^{(3)}(q_\odot, q, q_*) u\left(\lVert q - q_\odot\rVert_2\right) u\left(\lVert q - q_*\rVert_2\right) 
\end{align} 
\end{widetext}
These two surviving contributions are the off-diagonal terms labeled $C$ and $S$ in the main text (see Eqs.~(\ref{eq:C_term})–(\ref{eq:J_term})), corresponding respectively to the three-body and four-body correlation contributions.

\bibliography{apssamp}% Produces the bibliography via BibTeX.

\end{document}